\documentclass[journal, onecolumn, draftcls,12pt]{IEEEtran}
\usepackage{amssymb,amsmath,epsfig,graphicx,theorem}
\usepackage{amsfonts}
\usepackage{bm}
\usepackage{arydshln}
\usepackage[normalem]{ulem}
\usepackage{algorithm}
\usepackage{algorithmic}
\newtheorem{Theorem}{Theorem}
\newtheorem{Lemma}{Lemma}
\newtheorem{Corollary}{Corollary}

\newtheorem{Definition}{Definition}
\newtheorem{Remark}{Remark}

\usepackage{epsfig,rotating,setspace,latexsym,amsmath,epsf,amssymb,bm,color}
\usepackage{cite}
\usepackage{multirow}

\IEEEoverridecommandlockouts

\title{Multiple Threshold Schemes Under The Weak Secure Condition}

\author{Jiahong Wu, Nan Liu, Wei Kang%
\thanks{J. Wu and N. Liu are with the National Mobile Communications Research Laboratory,
Southeast University, Nanjing, China (email: \{jiahongwu,nanliu\}@seu.edu.cn). W. Kang is with the School of Information Science and Engineering,
Southeast University, Nanjing, China (email: wkang@seu.edu.cn). }%
\thanks{This work was partially supported by the National Natural Science Foundation of China
under Grants $61971135$, the Research Fund of National Mobile Communications Research Laboratory, Southeast University (No. 2020A03) and Six talent peaks project in Jiangsu Province.
}
}

\begin{document}

\maketitle
\begin{abstract}
In this paper, we consider the case that sharing many secrets among a set of participants using the threshold schemes. All secrets are assumed to be statistically independent and the weak secure condition is focused on. Under such circumstances we investigate the infimum of the (average) information ratio and the (average) randomness ratio for any structure pair which consists of the number of the participants and the threshold values of all secrets. 

For two structure pairs such that the two numbers of the participants are the same and the two arrays of threshold values have the subset relationship, two leading corollaries are proved following two directions. More specifically, the bound related to the lengths of shares, secrets and randomness for the complex structure pair can be truncated for the simple one; and the linear schemes for the simple structure pair can be combined independently to be a multiple threshold scheme for the complex one. The former corollary is useful for the converse part and the latter one is helpful for the achievability part.

Three new bounds special for the case that the number of secrets corresponding to the same threshold value $ t $ is lager than $ t $ and two novel linear schemes modified from the Vandermonde matrix for two similar cases are presented. Then come the optimal results for the average information ratio, the average randomness ratio and the randomness ratio. We introduce a tiny example to show that there exists another type of bound that may be crucial for the information ratio, to which we only give optimal results in three cases.
\end{abstract}


\section{introduction}


A secret sharing scheme is a method to share a secret among a set of participants such that only certain subsets of participants can recover the secret while any other subset obtains nothing\cite{shamir1979how}. 

Establishing bounds on the length of the shares to be given to participants in secret sharing schemes is one of the fundamental problems. If the length of the shares given to participants are too long compared to the length of the secret, the whole scheme may be inefficient. Researchers are interested in the (average) ratio between the length of the shares and the secret, the number of different structures is finite for fixed participant size $N$ and it turns out that linear schemes can achieve the best ratios when the number of participants $N\leq 4$ \cite{stinson1992an}. If $N=5$, \cite{jackson1996perfect} had already handled most structures. Until recently the work is moved further by introducing the converse for linear schemes and constructing the corresponding generator matrices \cite{9127978}. While the general results in converse or achievability are far from tight as discussed in \cite{padr2012lecture}.

The problem of estimating the amount of random bits necessary to set up the schemes has also received attention recently. As the amount of randomness needed by an algorithm is to be treated as a computational resource, analogously to the amount of time and space needed. \cite{blundo1996randomness} initiated a quantitative investigation of the number of random bits needed by secrets sharing schemes and found optimal results for some schemes. Some other results relate to this criteria can be found in \cite{farras2012linear,tang2014complexity}.

There are some situations where more than one secret is to be shared among participants. \cite{simmons1992introduction} introduced the missile battery scenario where not all of the missiles have the same launch enable code. Linear threshold multisecret sharing schemes are studied in \cite{farras2012linear} where the distributed secrets are in one-to-one correspondence with the sets of $ k $-out-of-$ N $ participants, i.e., the number of secrets equals $ \binom{N}{k} $ and the decoding condition relies on such set of length $ k $. Distributed multi-user secret sharing scheme is also emerged where secret sharing schemes are implied in the distributed storage systems \cite{khalesi2021capacity,soleymani2020distributed}.

From one secret to multiple secrets among the same participants, the decodable condition for the latter can be seen as considering multiple secrets individually. While the secure condition varies depending on two criteria: the strong version and the weak version. For example, the strong secure condition asks that any set of participants which is unable to decode secret $ S_1 $ or secret $ S_2 $ has no information on the whole set of these two secrets, however, the weak secure condition says that such set of participants has no information on any single one secret, which means some information about secrets $ S_1 $ and $ S_2 $ like linear combinations may be known. 

In \cite{ramp1999multiple}, the authors show that under the strong secure condition, when the secrets are statistically independent, the best we can do is combining individual ramp schemes independently. \cite{herranz2014new} has showed that under the weak secure condition, we may gain more efficiency at the cost of security. 

In this paper, we focus on the special model implied by \cite{ramp1999multiple}. More specifically, we involve threshold scheme such that any set of participant either decodes the secret or learns nothing, while ramp scheme is more general and has gaps. We follow the assumption that all secrets are statistically independent in order to obtain theoretical results that shining the fundamental limits like in \cite{ramp1999multiple}. The weak secure condition is focused on, as we care about multiple threshold scheme of any structure, in particular the case that the number of secrets corresponding to the same threshold value $ t $ is lager than $ t $, the existing different-threshold-bound in \cite{herranz2014new} is not enough. Like the (average) ratio and the randomness criteria discussed for single secret sharing scheme, we pursue the infimum of the (average) information ratio and the (average) randomness ratio of any possible structure for multiple threshold scheme and hope that such four ratios will give a basic understanding of this problem.

In this paper, every possible structure is of concern for us. For fixed number of participants, when two access structure arrays have the subset relationship, in the converse part we find that the region formed by the Shannon-type inequalities and system conditions of the simple structure is completely embedded in the region of the complex one; in the achievability part linear schemes of the simple structure can be combined independently to be a multiple threshold scheme for the complex structure. Thus the bridge between different structures is built and guaranteed theoretically.

The case that the number of secrets corresponding to the same threshold value $ t $ is lager than $ t $ is special and inspires three new bounds in addition to the different-threshold-bound in \cite{herranz2014new}, which are threshold-sum-difference-bound, threshold-product-bound and threshold-sum-bound. Existing linear scheme when under the weak secure condition is based on the Vandermonde matrix as in \cite{mceliece1981sharing,karnin1983secret,herranz2014new,csirmaz2011share}. Still in this case we propose two novel linear schemes which are modified from the Vandermonde matrix like in \cite{khalesi2021capacity} to meet the bounds. And we can learn that as the weak secure condition asks the protection of any single one secret, the assumption that the secrets are statistically independent may promote this requirement.

Finally relate these new findings and existing results to the four ratios, we claim optimality in the average information ratio, the average randomness ratio and the randomness ratio. Still new patterns of bounds need to be investigated further as we only get optimal results for the information ratio in three cases.


\section{problem formulation} \label{sec_p_f}
\subsection{system model}\label{sys_model}
A $ (t, N, S) $ threshold scheme is a protocol to distribute a secret $ S $ among a set of $ N $ participants in such a way that any set of participants of  cardinality greater than or equal to $ t(t\leq N) $ can decode the secret $ S $ and any set of participants of cardinality less than or equal to $ t-1(t\geq2) $ has no information on $ S $, we refer these two conditions as the decodable condition and the secure condition.

Such protocol is treated as a special discrete probability distribution with $ N+1 $ random variables \cite{stinson1992an}: the secret $ S $ and the $ i $-th share $ P_i $ of the $ i $-th participant for every $ i $ in the set $ [N]:=\{1,2,\ldots,N\}(N\geq2) $. In this way suppose a \textit{dealer} wants to share the secret $ S=s $ among the $ N $ participants, he does this by giving the $ i $-th participant the share $ P_i=p_i $ for every $ i\in[N] $ according to the conditional probability $ \text{Pr}(P_1=p_1,\ldots,P_N=p_N|S=s) $, where $ \{s,p_1,\ldots,p_N\} $ are chosen from a finite field. The specificity of this discrete probability distribution corresponds to the two requirements of the protocol. More specifically, given any set of shares of cardinality greater than or equal to $ t $, the conditional probability of the secret $ S=s $ is equal to 1 or 0, i.e., the conditional entropy of the secret $ S $ given $ t $ or more shares equals 0; given any
set of shares of cardinality less than or equal to $ t-1 $, the conditional probability of the secret $ S=s $ is equal to $ \text{Pr}(S=s) $, i.e., the conditional entropy of the secret $ S $ given $ t-1 $ or less shares equals the entropy of the secret $ S $, in other words, the secret $ S $ and $ t-1 $ or less shares are statistically independent.

In this paper we consider the case in which we want to share many secrets among a set of $ N $ participants, using the threshold schemes. As every single one secret has its corresponding threshold value $ t $ determining the decodable and secure conditions, we naturally arrange the threshold values of all secrets in \emph{non-increasing order} without loss of generality in an array $ \mathcal{T} $, named as the access structure array. 

From one secret to multiple secrets among the same $ N $ participants, the decodable condition for the latter can be seen as considering multiple secrets individually, e.g., any set of participants of cardinality greater than or equal to the first element of  the access structure array can decode all secrets due to the non-increasing order. While the secure condition varies depending on two criteria: the strong version and the weak version. For example, the strong secure condition asks that any set of participants of cardinality strictly less than the last element of  the access structure array has no information on the whole set of $ |\mathcal{T}| $ secrets, however, the weak secure condition says that such set of participants has no information on any single one secret, which means some information about all $ |\mathcal{T}| $ secrets like linear combinations may be known. Still use the technique of discrete probability distribution, we define a multiple threshold scheme as follows.
\begin{Definition}[Multiple Threshold Scheme]\normalfont
	The structure of a multiple threshold scheme is denoted by the pair $ (N,\mathcal{T}=(T_1,\ldots,T_K)) $, where $ N $ is the number of participants, $ K $ is the number of unique elements in the access structure array $ \mathcal{T} $ since there may be some secrets with the same threshold value, and for any $ k\in[K] $, $ T_k $ is also an array with length $ |T_k| $ full of $ t_k $, the same threshold value of the corresponding set of secrets $ \mathcal{S}_k:=\{S_{k,1},\ldots,S_{k,|T_k|}\} $. An $ (N,\mathcal{T}) $ multiple threshold scheme is a discrete probability distribution with $ N+|\mathcal{T}| $ random variables of which the $ |\mathcal{T}| $ secrets are assumed to be statistically independent, i.e.,
	\begin{equation}\label{secrets_independent}
		H(\mathcal{S}_1,\ldots,\mathcal{S}_K)=\sum_{k\in[K]}\sum_{j\in[|T_k|]}H(S_{k,j}).
	\end{equation}

	Like threshold scheme, the decodable and secure conditions are made in the following. For any $ k\in[K] $,
	\begin{enumerate}
		\item The decodable condition: Any set of at least $ t_k $ participants can decode the set $ \{\mathcal{S}_k,\ldots,\mathcal{S}_K\} $ of secrets. Formally, for all $ A\subseteq[N] $ with $ |A|\geq t_k $, let $ P_A:=\{P_i|i\in A\} $, it holds that
		\begin{equation}\label{decodable_condition}
			H(\mathcal{S}_k,\ldots,\mathcal{S}_K |P_A)=0.
		\end{equation}
	\item The strong secure condition: Any set of at most $ t_k-1 $ participants has no information on the set $ \{\mathcal{S}_1,\ldots,\mathcal{S}_k\} $ of secrets. Formally, for all $ A\subseteq[N] $ with $ |A|\leq t_k-1 $, it holds that
	\begin{equation}\label{strong_condition}
		H(\mathcal{S}_1,\ldots,\mathcal{S}_k|P_A)=H(\mathcal{S}_1,\ldots,\mathcal{S}_k).
	\end{equation}
\item The weak secure condition: Any set of at most $ t_k-1 $ participants has no information on any single one secret $ S_{i,j} $, where $ i\in[k] $ and $ j\in[|T_i|]  $. Formally, for all $ A\subseteq[N] $ with $ |A|\leq t_k-1 $, it holds that
\begin{equation}\label{weak_condition}
	H(S_{i,j}|P_A)=H(S_{i,j}).
\end{equation}
	\end{enumerate}
\end{Definition}
\begin{Remark}\normalfont
	Note that these two secure conditions are different and each will be investigated when combined with the same decodable condition and the assumption that all secrets are statistically independent. When the length of the access structure array $ \mathcal{T} $  equals $ 1 $, the assumption of secrets is meaningless, both strong and weak secure condition are the same, that is, the system conditions of multiple threshold scheme degrade to those of threshold scheme.
\end{Remark}

The efficiency of a threshold scheme is usually measured by the ratio between the length of a share and the secret. Form a set of entropy of every single share, the information ration is defined as the maximum value of this set divided by the entropy of the secret and the average information ratio is defined similarly except that the numerator is replaced by the average value of the same set. As for an $ (N,\mathcal{T}) $ multiple threshold scheme, the information ration $ \sigma $ and the average information ration $ \tilde{\sigma} $ are defined similarly as follows:
\begin{align}
	&\sigma=\frac{\max_{i\in[N]}H(P_i)}{\min_{k\in[K],j\in[|T_k|]}H(S_{k,j})},\\	
	&\tilde{\sigma}=\frac{\sum_{i\in[N]}H(P_i)/N}{\sum_{k\in[K]}\sum_{j\in[|T_k|]}H(S_{k,j})/|\mathcal{T}|}.
\end{align}
Note that the major difference for a multiple threshold scheme is that the denominator of the information ratio is replaced by the minimum of the length of every secret as in \cite{farras2012linear} and the average information ratio considers the average length of all secrets to divide.

For fixed $ N $ participants, since a multiple threshold scheme has more random variables than a threshold scheme, the former is more complicated. Then comes the notion of randomness, a fundamental computation resource because truly random bits are hard to obtain. The total randomness present in an $ (N,\mathcal{T}) $ multiple threshold scheme is the entropy of the whole $ N $ shares, i.e., $ H(P_{[N]}) $, which takes into account also the randomness $ H(\mathcal{S}_{[K]}) $ of all secrets as $ N $ shares can decode all secrets. In this paper we mainly consider the randomness for the dealer to set up a multiple threshold scheme for secrets $ \mathcal{S}_{[K]} $, that is, the randomness he uses to generate the $ N $ shares given the marginal discrete probability distribution of secrets $ \mathcal{S}_{[K]} $. Therefore, we define the randomness ratio $ \tau $ as the difference between the length of $ N $ shares and that of $ |\mathcal{T}| $ secrets, then divided by the minimum of the set formed by the length of every secret. The definition for the average randomness ratio is similar except that the denominator is replaced by the average value of such set. More specifically,
\begin{align}
	&\tau=\frac{H(P_{[N]})-H(\mathcal{S}_{[K]})}{\min_{k\in[K],j\in[|T_k|]}H(S_{k,j})},\\	
	&\tilde{\tau}=\frac{H(P_{[N]})-H(\mathcal{S}_{[K]})}{\sum_{k\in[K]}\sum_{j\in[|T_k|]}H(S_{k,j})/|\mathcal{T}|}.
\end{align}
As all secrets are statistically independent by the assumption, the randomness of all secrets can be replaced by the sum of entropy of every single secret as in equation \eqref{secrets_independent}.

The problem that we investigate in this paper is to optimize the (average) information ratio and the (average) randomness ratio of multiple threshold schemes. Given the number of participants $ N $ and the access structure array $ \mathcal{T} $, we define $ (\tilde{\sigma}_{N,\mathcal{T}})\sigma_{N,\mathcal{T}} $ as the infimum of the (average) information ration of all $ (N,\mathcal{T}) $ multiple threshold schemes, moreover, $ \tilde{\tau}_{N,\mathcal{T}} $ and $ \tau_{N,\mathcal{T}} $ are defined similarly. For all possible choices of the structure $ (N,\mathcal{T}) $, we are interested in determining the above four values via their lower bounds and upper bounds.

\subsection{lower bound}
Note that for any probability distribution with $ N+|\mathcal{T}| $ discrete random variables, we can extract its $ 2^{N+|\mathcal{T}|}-1 $ entropies and arrange them into a vector $ \mathbf{h} $. Any such vector $ \mathbf{h} $ satisfies the so-called Shannon-type inequalities \cite{zhang1998on}. More specifically,  let $\mathcal{H}_{N+|\mathcal{T}|}$ be a $(2^{N+|\mathcal{T}|}-1)$-dimension Euclidean space whose coordinates are labeled by $h_a$, $\emptyset\neq a\subseteq \mathcal{O}$, where $ \mathcal{O} $ is the set of $ N+|\mathcal{T}| $ random variables $ \{P_i:i\in[N]\}\cup\{S_j:j\in[|\mathcal{T}|]\} $ and we use another notation for the secrets. Then let $\Gamma_{N+|\mathcal{T}|}$ be a polyhedral cone represented by the intersection of two categories of closed half-spaces, which are derived from the Shannon-type inequalities:
\begin{enumerate}
  \item Non-decreasing: If $a\subseteq b\subseteq\mathcal{O}$, then $h_a\leq h_b$,
  \item Submodular: $\forall a,b\subseteq\mathcal{O},h_{a\cup b}+h_{a\cap b}\leq h_a+h_b$,
\end{enumerate}
where $ h_\emptyset $ is taken to be 0. Note that for every discrete probability distribution and the corresponding vector $ \mathbf{h} $, we have (conditional) entropy and (conditional) mutual information greater than and equal to 0, which correspond to being non-decreasing and submodular respectively, then $ \mathbf{h}\in\Gamma_{N+|\mathcal{T}|} $.

As the independent assumption for secrets \eqref{secrets_independent}, the decodable condition \eqref{decodable_condition}, the strong secure condition \eqref{strong_condition} and the weak secure condition \eqref{weak_condition} can all be handled as homogeneous linear equations involving the coordinates from $\mathcal{H}_{N+|\mathcal{T}|}$ only, for fixed structure $ (N,\mathcal{T}) $ of any multiple threshold scheme, we add four intersections of  hyperplane(s) as follows. Let $ [k:K]:=\{k,\ldots,K\} $, for every $ k\in[K] $ and $ A\subseteq[N] $,
\begin{align}
	 \mathcal{C}_0=&\{\mathbf{h}\in\mathcal{H}_{N+|\mathcal{T}|}:h_{\mathcal{S}_{[K]}}-\sum_{i\in[K]}\sum_{j\in|T_i|}h_{S_{i,j}}=0\},\\
  \mathcal{C}_1=&\{\mathbf{h}\in\mathcal{H}_{N+|\mathcal{T}|}:h_{\mathcal{S}_{[k:K]},P_A}-h_{P_A}=0,|A|\geq t_k\},\\
  \mathcal{C}_2=&\{\mathbf{h}\in\mathcal{H}_{N+|\mathcal{T}|}:h_{\mathcal{S}_{[k]},P_A}-h_{P_A}-h_{\mathcal{S}_{[k]}}=0,|A|\leq t_k-1\},\\
   \mathcal{C}_3=&\{\mathbf{h}\in\mathcal{H}_{N+|\mathcal{T}|}:h_{S_{i,j},P_A}-h_{P_A}-h_{S_{i,j}}=0,|A|\leq t_k-1,i\in[k],j\in[|T_i|]\}.
\end{align}
In this way, consider a vector $\mathbf{h}\in\mathcal{H}_{N+|\mathcal{T}|}$ whose components are the $ 2^{N+|\mathcal{T}|}-1 $ entropies obtained from an $ (N,\mathcal{T}) $ multiple threshold scheme under weak secure condition \eqref{weak_condition}, we have $\mathbf{h}\in\Gamma_{N+|\mathcal{T}|}\cap\mathcal{C}_{0,1,3}$, here we denote $ \mathcal{C}_0\cap\mathcal{C}_1\cap\mathcal{C}_3 $ by $ \mathcal{C}_{0,1,3} $ for ease of notation. And when replace weak secure condition by strong one, it follows that the corresponding vector belongs to $\Gamma_{N+|\mathcal{T}|}\cap\mathcal{C}_{0,1,2}$.

For a fixed structure $ (N,\mathcal{T}) $, recall the definitions of the four ratios above, besides their connections with any discrete probability distributions meeting some system conditions, there are also some quantities of interest, i.e., the set of $ N+|\mathcal{T}| $ entropies of every share and secret $ \{H(P_i):i\in[N]\}\cup\{H(S_j):j\in[|\mathcal{T}|]\} $ for the (average) information ratio and the set of $ 1+|\mathcal{T}| $ entropies of $ N $ shares and every secret $ \{H(P_{[N]})\}\cup\{H(S_j):j\in[|\mathcal{T}|]\} $ for the (average) randomness ratio. As for the polyhedral cone $ \Gamma_{N+|\mathcal{T}|}\cap\mathcal{C}_{0,1,(2)3} $ under (strong) weak secure condition, of which we care about the sets of the corresponding coordinates likewise, i.e., $ \mathbf{h}_I:=\{h_{P_i}:i\in[N]\}\cup\{h_{S_j}:j\in[|\mathcal{T}|]\} $ for the (average) information ratio and $ \mathbf{h}_R:=\{h_{P_{[N]}}\}\cup\{h_{S_j}:j\in[|\mathcal{T}|]\} $ for the (average) randomness ratio. In this way, the region involving these coordinates is enough to derive the lower bounds for the four ratios. To gain a more exact characterization, a suitable concept is illustrated as follows: A \textit{projection} \cite{kaluzny2002polyhedral} of a region $ \mathcal{P} $ in $ \mathbb{R}^n=\mathbb{R}^{n_1}\times\mathbb{R}^{n_2} $ onto its subspace of the first $ n_1 $ coordinates is
\begin{equation}\label{def_proj}
	\textrm{proj}_{[n_1]}(\mathcal{P})=\{\mathbf{x}_1\in\mathbb{R}^{n_1}:\exists\mathbf{x}_2,(\mathbf{x}_1^T,\mathbf{x}_2^T)^T\in\mathcal{P}\}.
\end{equation}
Then $ \text{proj}_{\mathbf{h}_I}(\Gamma_{N+|\mathcal{T}|}\cap\mathcal{C}_{0,1,(2)3}) $ and $ \text{proj}_{\mathbf{h}_R}(\Gamma_{N+|\mathcal{T}|}\cap\mathcal{C}_{0,1,(2)3}) $ are essential for the (average) information ratio and (average) randomness ratio respectively.

Recall that every possible structure $ (N,\mathcal{T}) $ is of concern for us. For fixed number $ N $ of participants, two different access structure arrays $ \mathcal{T}'=(T_1',\ldots,T_{K'}') $ and $ \mathcal{T}=(T_1,\ldots,T_{K}) $ may have a relationship similar to the subset concept from set theory, thus we give the following definition with some abuse of notation:
\begin{Definition}[Subset of Arrays]\label{definition_subset_array}
	$ \mathcal{T}'\subseteq\mathcal{T} $ if and only if for every $ i\in[K'] $, there exists $ j\in[K] $ such that $ t_i'=t_j $ and $ |T_i'|\leq|T_j| $, where $ t_i' $ is the same threshold value of the sub-array $ T_i' $. 
\end{Definition}

When $ \mathcal{T}'\subseteq\mathcal{T} $ and for the same number $ N $ of participants, in the polyhedral cone formed by the Shannon-type inequalities, there exist some interesting geometrical properties between two regions $ \Gamma_{N+|\mathcal{T}|}\cap\mathcal{C}_{0,1,(2)3} $ and $ \Gamma_{N+|\mathcal{T}'|}\cap\mathcal{C}_{0,1,(2)3}' $, where $ \mathcal{C}_.' $ is the intersection of hyperplanes defined from the structure $ (N,\mathcal{T}') $. For the ease of notation we reuse  coordinates of $ \Gamma_{N+|\mathcal{T}'|}\cap\mathcal{C}_{0,1,(2)3}' $ from $ \Gamma_{N+|\mathcal{T}|}\cap\mathcal{C}_{0,1,(2)3} $ and rearrange the elements of these two access structure arrays such that for every $ i\in[|\mathcal{T}'|] $, the two corresponding secrets from $ \mathcal{T}' $ and $ \mathcal{T} $ respectively have the same threshold value, here the non-increasing order of both access structure arrays may be broken due to such forcible arrangement. Denote all coordinates of the former region by $ \mathbf{h}_L $ and we give the following theorem:
\begin{Theorem}\label{subset_region_equal}
	When the number of participants $ N $ is fixed and two access structure arrays have the subset relationship, i.e., $ \mathcal{T}'\subseteq\mathcal{T} $, the polyhedral cone formed by the Shannon-type inequalities and system conditions of the structure $ (N,\mathcal{T}') $ is exactly equal to the projection of the corresponding polyhedral cone of the structure $ (N,\mathcal{T}) $, i.e., $ \Gamma_{N+|\mathcal{T}'|}\cap\mathcal{C}_{0,1,(2)3}'=\text{proj}_{\mathbf{h}_L}(\Gamma_{N+|\mathcal{T}|}\cap\mathcal{C}_{0,1,(2)3}) $.
\end{Theorem}
\begin{IEEEproof}
	The proof is conducted in two directions, i.e., any point of the former is in the later, and vice versa. The detail is put in Appendix \ref{proof_theorem_1}.
\end{IEEEproof}
\begin{Remark}\normalfont
	This theorem offers a theoretical guarantee that some important bounds of the bigger region, after a projection algorithm like Fourier-Motzkin-Elimination, are still feasible and even fundamental for the smaller region. A direct example is the following corollary.
\end{Remark}
\begin{Corollary}\label{bound_complex_simple}
	When the number of participants $ N $ is fixed and two access structure arrays have the subset relationship, i.e., $ \mathcal{T}'\subseteq\mathcal{T} $, a feasible bound for the polyhedral cone formed by the Shannon-type inequalities and system conditions of the structure $ (N,\mathcal{T}) $ can be truncated to be a feasible bound of the corresponding polyhedral cone of the structure $ (N,\mathcal{T}') $. More specifically, for any vector $ \mathbf{h}\in\Gamma_{N+|\mathcal{T}|}\cap\mathcal{C}_{0,1,(2)3} $, if there exists a feasible bound such that $ \alpha_0h_{P_{[N]}}+\sum_{i\in[N]}\alpha_ih_{P_i}\geq\sum_{j\in[|\mathcal{T}|]}\beta_jh_{S_j} $, where every coefficient $ \beta_j $ is non-negative, then for any vector $ \mathbf{h}'\in\Gamma_{N+|\mathcal{T}'|}\cap\mathcal{C}_{0,1,(2)3}' $, $ \alpha_0h_{P_{[N]}}'+\sum_{i\in[N]}\alpha_ih_{P_i}'\geq\sum_{j\in[|\mathcal{T}'|]}\beta_jh_{S_j}' $.
\end{Corollary}
\begin{IEEEproof}
	For any $ i\in[|\mathcal{T}|]$, consider a submodular inequality $ h_{S_i}+h_{P_1}-h_{S_i,P_1}\geq0 $ and a non-decreasing one $ -h_{P_1}+h_{S_i,p_1}\geq0 $, plus these two together, we get $ h_{S_i}\geq 0 $. Then $ \alpha_0h_{P_{[N]}}+\sum_{i\in[N]}\alpha_ih_{P_i}\geq\sum_{j\in[|\mathcal{T}'|]}\beta_jh_{S_j} $ is also a feasible bound. Based on the equation result of theorem \ref{subset_region_equal}, such bound is also feasible for the vectors of the region $ \Gamma_{N+|\mathcal{T}'|}\cap\mathcal{C}_{0,1,(2)3}' $.
\end{IEEEproof}
\begin{Remark}\normalfont
	For an access structure array with continuous unique threshold values, the proofs of some bounds may be more read-friendly. And by this corollary, the bound for a non-continuous access structure array $ \mathcal{T}' $ can be obtained by the truncation from the bound of a continuous access structure array $ \mathcal{T}\supseteq\mathcal{T}'  $, and it turns out that some bounds obtained in this way are enough for the four ratios.
\end{Remark}

\subsection{linear scheme}
Before introducing the so-called \textit{linear scheme} \cite{karnin1983secret}, some concepts need to be illustrated.  Let $\mathcal{V}$ be a vector space with finite dimension $n$ over a finite field $\mathbb{F}_q$, where $q$ is a  prime number. Recall that $ \mathcal{O} $ is the set of $ N+|\mathcal{T}| $ random variables essential to an $ (N,\mathcal{T}) $ multiple threshold scheme. For every $i\in\mathcal{O}$, denote a subset of $\mathcal{V}$ by ${V}_i$, we assume that all vectors of ${V}_i$ are linear independent, then the length of such subset is $ \text{rk}(V_i) $, here $ \text{rk}(.) $ calculates the rank of its input. Hereafter, we treat each ${V}_i$ as a matrix of shape $ n\times\text{rk}(V_i)$ and stack such $ |\mathcal{O}| $ matrices in sequence horizontally (column wise) into a bigger matrix $\mathbf{G}\in \mathbb{F}_q^{n\times\sum_{i\in\mathcal{O}}\text{rk}({V}_i)}$, referred as a generator matrix. 

Then consider a uniform discrete probability distribution whose alphabet is the set of $q^n$ different row vectors with length $n$ taking values from $\mathbb{F}_q$. These vectors can be stacked vertically into a matrix $\mathbf{C}\in \mathbb{F}_q^{q^n\times n}$. For $\mathbf{CG}\in \mathbb{F}_q^{q^n\times \sum_{i\in\mathcal{O}}\text{rk}({V}_i)}$, with the set of row vectors as the alphabet and that the probability of each equals $1/q^n$, it forms a new discrete probability distribution with random variables from the set $\mathcal{O}$, where $\forall\emptyset\neq x\subseteq\mathcal{O},H(x)=\text{rk}({V}_x):=\text{rk}(\cup_{i\in x} {V}_i)$, with the base of the logarithm being $ q $ and $ {V}_x $ denotes a matrix stacked horizontally by the corresponding $ |x| $ matrices. Note that if $\text{rk}({V}_{\mathcal{O}})<n$, then some row vectors of $\mathbf{CG}$ are the same and the corresponding probabilities need to be summed; otherwise, all are unique.

Recall that $ K $ is the number of unique elements in the access structure array $ \mathcal{T} $ and for any $ k\in[K] $, $ T_k $ is also an array with length $ |T_k| $ full of $ t_k $, the same threshold value of the corresponding set of secrets $ \mathcal{S}_k=\{S_{k,1},\ldots,S_{k,|T_k|}\} $. Based on the relationship between discrete probability distribution and generator matrix, consider the latter as a linear scheme, then we give the following definition of linear multiple threshold scheme:
\begin{Definition}[Linear Multiple Threshold Scheme]\label{linear_mts}
	A linear $ (N,\mathcal{T}) $ multiple threshold scheme is a special generator matrix satisfying the following system conditions:
	\begin{enumerate}
		\item All secrets are statistically independent: $\text{rk}(V_{\mathcal{S}_{[K]}})=\sum_{k\in[K]}\sum_{j\in[|T_k|]}\text{rk}(V_{S_{k,j}})$.
		\item The decodable condition: For every $ k\in[K]$ and all $ A\subseteq[N] $ with $ |A|\geq t_k $, $\text{rk}(V_{\mathcal{S}_{[k:K]},P_A})=\text{rk}(V_{P_A})$.
		\item The strong secure condition: For every $ k\in[K]$ and all $ A\subseteq[N] $ with $ |A|\leq t_k-1 $, $\text{rk}(V_{\mathcal{S}_{[k]},P_A})=\text{rk}(V_{\mathcal{S}_{[k]}})+\text{rk}(V_{P_A})$.	
		\item The weak secure condition: For every $ k\in[K]$ and all $ A\subseteq[N] $ with $ |A|\leq t_k-1 $, $\text{rk}(V_{{S}_{i,j},P_A})=\text{rk}(V_{{S}_{i,j}})+\text{rk}(V_{P_A})$, where $ i\in[k] $ and $ j\in[|T_i|] $.
	\end{enumerate}
\end{Definition}
	Note that these four conditions are rewritten from \eqref{secrets_independent}, \eqref{decodable_condition}, \eqref{strong_condition} and \eqref{weak_condition} in section \ref{sys_model}, still only one of the two secure conditions is chosen for a corresponding model and for simplicity we do not mention which secure condition is implied in this subsection.
\begin{Remark}\normalfont
	From the view of a linear $ (N,\mathcal{T}) $ multiple threshold scheme, both the generation of $ N $ shares and the decoding of secrets can be treated as linear mappings, which are more neat than the general ones.  As every codeword corresponds to $ |\mathcal{T}| $ secret values and a distribution of shares, furthermore, the whole probability distribution is uniform, then the length of every share is exactly the number of symbols given to the corresponding participant from the dealer. 
	The decodable condition tells that every vector $ \mathbf{a} $ in $V_{\mathcal{S}_{[k:K]}}$ can be written as a linear combination of vectors in $V_{P_A}$, which means that any $ t_k $ or more shares determine the secrets $ \mathcal{S}_{[k:K]} $ linearly.
	As for the (strong) weak secure condition, all elements of the union of $V_{(\mathcal{S}_{[k]})S_{i,j}}$ and $V_{P_A}$ are linear independent, which implies that the random variables $(\mathcal{S}_{[k]})S_{i,j}$ and $P_A$ constructed are statistically independent. The independent assumption for all secrets follows similarly.
\end{Remark}

In this paper we only use linear multiple threshold schemes for achievability. Then the entropy part from all four ratios can be replaced by the the rank part. For a fixed structure $ (N,\mathcal{T}) $, if we can construct a corresponding linear multiple threshold scheme, i.e., a generator matrix satisfying the system conditions, we have upper bounds of the four ratios. When upper bound and lower bound match, the optimum can be claimed. 

As we care about every possible structure $ (N,\mathcal{T}) $, recall that for fixed number of participants $ N $, two different access structure arrays $ \mathcal{T}'=(T_1',\ldots,T_{K'}') $ and $ \mathcal{T}=(T_1,\ldots,T_{K}) $ may have a subset relationship similar to two sets according to definition \ref{definition_subset_array}. Still for the ease of notation, we rearrange the elements of these two access structure arrays such that for every $ i\in[|\mathcal{T}'|] $, the two corresponding secrets from $ \mathcal{T}' $ and $ \mathcal{T} $ respectively have the same threshold value, finally we claim that
\begin{Theorem}\label{simple_point_complex}
	When the number of participants $ N $ is fixed and two access structure arrays have the subset relationship, i.e., $ \mathcal{T}'\subseteq\mathcal{T} $, any generator matrix $ V $ for the structure $ (N,\mathcal{T}') $ is also a linear $ (N,\mathcal{T}) $ multiple threshold scheme which reuses the random variable of the former and $ \text{rk}(V_{S_i})=0,\forall i\in[|\mathcal{T}'|+1,|\mathcal{T}|] $.
\end{Theorem}
\begin{IEEEproof}
	The proof is carried by checking the system conditions of the definition of linear multiple threshold scheme. The detail can be found in Appendix \ref{proof_theorem_2}.
\end{IEEEproof}
\begin{Remark}\normalfont
	The construction of a linear multiple threshold scheme may be easier when there are not too many secrets to share, i.e., the length of the access structure array is small. In this way, this theorem offers a degradation method to construct a more complicated scheme. As will be discussed below, it turns out to be a building block from simple schemes to a complex one.
\end{Remark}

Different generator matrices can be combined independently, i.e., in the diagonal, to form a bigger generator matrix  of which every rank term equals the sum of those from the original generator matrices. The detail is in the following:
\begin{Definition}[Independent Combination of Generator Matrices]\label{independent_combination}
	For the same set of random variables $ \mathcal{O} $, assume over the same finite field there are $ A $ existing generator matrices, each is denoted by $ V^i,\ i\in[A] $ and the corresponding shape is $ \text{rk}(V_\mathcal{O}^i)\times\sum_{j\in\mathcal{O}}\text{rk}(V_j^i) $. The independent combination of these $ A $ generator matrices is a bigger generator matrix $ V $, where for any $ j\in\mathcal{O} $ the sub-matrix $ V_j $ is a diagonal arrangement of the corresponding $ A $ sub-matrices from the original generator matrices. More specifically,  let the shape of the sub-matrix $ V_j $ be $ \sum_{i\in[A]}\text{rk}(V_\mathcal{O}^i) \times\sum_{i\in[A]}\text{rk}(V_j^i)$ and fill $ V_j $ with zeroes, then for every $ i\in[A] $ in turn put $ V_j^i $ in the diagonal of $ V_j $.
\end{Definition}
Note that the we can do Gaussian-Elimination towards the rows of a generator matrix and the corresponding $ 2^{|\mathcal{O}|}-1 $ rank terms do not change, in this way we fix the number of rows of a generator matrix to be the rank of the whole matrix. When there exist an all-zero $ V_j^i $, we do nothing for the sub-matrix $ V_j $ of the final bigger generator matrix when it is the turn of the sub-matrix $ V_j^i $ from original generator matrices in the diagonal arrangement.
\begin{Remark}\normalfont
	By the diagonal arrangement and based on the same finite field, it follows that for any non-empty set $ B\subseteq\mathcal{O} $
	\begin{equation}\label{independent_rk_sum_equation}
		\text{rk}(V_B)=\sum_{i\in[A]}\text{rk}(V_B^i).
	\end{equation}
The rigorous proof can follow the direct implementation of Gaussian-Elimination horizontally (column wise) and is omitted here.
\end{Remark}

Via the claim that a simple linear multiple threshold scheme can be treated as a complex one with somewhat of degradation in theorem \ref{simple_point_complex}, and using the construction method that independently combing existing linear multiple threshold schemes as in definition \ref{independent_combination}, we give the following corollary.
\begin{Corollary}\label{point_simple_complex}
	When the number of participants $ N $ is fixed and $ A+1 $ access structure arrays have the subset relationship, i.e., $ \mathcal{T}^i\subseteq\mathcal{T},\ i\in[A]$. Assume over the same finite field there are $ A $ existing generator matrices, each is denoted by $ V^i $ and for the structure $ (N,\mathcal{T}^i) $. Then the independent combination of these $ A $ generator matrices is a linear $ (N,\mathcal{T}) $ multiple threshold scheme.
\end{Corollary}
\begin{IEEEproof}
	From theorem \ref{simple_point_complex}, for any $ i\in[A] $, a linear scheme for the structure $ (N,\mathcal{T}^i) $ is also a linear $ (N,\mathcal{T}) $ multiple threshold scheme when $ \mathcal{T}^i\subseteq\mathcal{T} $. Recall the system conditions of a linear multiple threshold scheme in definition \ref{linear_mts}, each of them is some equations involving the rank terms only. As the independent combination results in the sum of rank terms as in \eqref{independent_rk_sum_equation}, the conclusion follows.
\end{IEEEproof}
\begin{Remark}\normalfont
From one secret to multiple ones, a trivial idea to accomplish the construction of an $ (N,\mathcal{T}) $ multiple threshold scheme is combining $ |\mathcal{T}| $ threshold schemes independently. As for liner schemes, this corollary applies the notion of independence and builds a bridge from some simple schemes to a complex one. It turns out that some linear schemes obtained in this way are enough for the four ratios.
\end{Remark}

\section{(Average) information ratio}\label{ir}
We firstly discuss multiple threshold schemes with  the strong secure condition, of which some results can be modified for the weak one.
\subsection{strong secure}\label{strong_secure_ir}
\subsubsection{converse}
We firstly discuss some converse results that are fundamental for lower bounds of the (average) information ratio.

\begin{Lemma}[Size of A Secret]\label{lemma_secret}
	For any $ (t, N, S) $ threshold scheme, if $ t<N $, then $$ H(S)\leq I(P_{i_1};P_{i_2}|P_{i_3},\ldots,P_{i_{t+1}}), $$ where $ (i_1,\ldots,i_{t+1})$ are $ t+1 $ different values chosen from the set $ [N] $.
\end{Lemma}
\begin{IEEEproof}
To improve readability, we replace the index $ t_i $ by $ i $, the final result can be transformed back later. Hereafter, we use this trick in the following whole paper.
	\begin{align}
	H(S)=&H(S|P_{3},\ldots,P_{t+1})-H(S|P_{1},P_{3},\ldots,P_{t+1})\nonumber\\
	&-H(S|P_{2},P_{3},\ldots,P_{t+1})+H(S|P_{1},\ldots,P_{t+1})\nonumber\\
	=&I(S;P_{1}|P_{3},\ldots,P_{t+1})-I(S;P_{1}|P_{2},P_{3},\ldots,P_{t+1})\nonumber\\
	=&H(P_{1}|P_{3},\ldots,P_{t+1})-H(P_{1}|S,P_{3},\ldots,P_{t+1})\nonumber\\
	&-H(P_{1}|P_{2},P_{3},\ldots,P_{t+1})+H(P_{1}|S,P_{2},P_{3},\ldots,P_{t+1})\nonumber\\
	=&I(P_{1};P_{2}|P_{3},\ldots,P_{t+1})-I(P_{1};P_{2}|S,P_{3},\ldots,P_{t+1})\nonumber\\
	\leq &I(P_{1};P_{2}|P_{3},\ldots,P_{t+1})\nonumber
\end{align}
Note that the first equation is nothing but the decodable and secure conditions of a $ (t, N, S) $ threshold scheme, the other equations are from the definitions of (conditional) entropy and (conditional) mutual information, and the last inequality uses the submodular property.
\end{IEEEproof}
\begin{Remark}\normalfont
	Existing result shows that the length of a share must be bigger than or equal to the length of a secret  \cite{padr2012lecture}, for example, if $ t=N $, $ H(S)=H(S|P_{i_2},\ldots,P_{i_{t}})-H(S|P_{i_1},\ldots,P_{i_{t}})=I(S;P_{i_1}|P_{i_2},\ldots,P_{i_{t}})\leq H(P_{i_1}|P_{i_2},\ldots,P_{i_{t}})\leq H(P_{i_1}) $. While this lemma tells that the length of a secret is not larger than the mutual information between any two shares given any other $ t-1 $ shares. In this sense, this bound is tighter and will be a building block for other interesting bounds of this paper.
\end{Remark}

From one secret to multiple ones, when under the strong secure condition, the following bound is enough for the investigation of the (average) information ratio and even the heterogeneous-secret-size problem as in \cite{ramp1999multiple}.
\begin{Lemma}\label{strong_sum_bound}
	For any $ (N,\mathcal{T}) $ multiple threshold scheme with the strong secure condition, it holds that $ \sum_{k\in[K]}\sum_{j\in[|T_k|]}H(S_{k,j})\leq H(P_i) $, $ \forall i\in[N] $.
\end{Lemma}
\begin{IEEEproof}
	The proof is based on the mutual information bound as in lemma \ref{lemma_secret}. The detail is in Appendix \ref{proof_strong_sum_bound}.
\end{IEEEproof}
\begin{Remark}\normalfont
	This bound says that when under the strong secure condition, the sum of the length of every secret is no bigger than the length of any single share. This is implied by the work from \cite{ramp1999multiple}, while our proof may be more read-friendly due to the building block in lemma \ref{lemma_secret}. Then consider the (average) information ratio, for any $ (N,\mathcal{T}) $ multiple threshold scheme, we apply this bound and it turns out that $ \sigma_{N,\mathcal{T}}\geq|\mathcal{T}| $ and $ \tilde{\sigma}_{N,\mathcal{T}}\geq|\mathcal{T}| $, i.e., lower bounds are formed.
\end{Remark}

\subsubsection{achievability}
Then we discuss some linear schemes that are crucial for upper bounds of the (average) information ratio.

Still we want to link the known results of a single secret to multiple ones. For a linear $ (t, N, S) $ threshold scheme, its generator matrix is closely related to the Vandermonde matrix over a finite field. More specifically, let $ V(t,[N+1]) $ fully denote the following matrix from $ \mathbb{F}_q^{t\times (N+1)} $, where $ q $ is a prime number larger than $ N+1 $. In this notation, $ t $ and $ [N+1] $ means the number of rows of the matrix and the elements of the second row in increasing order respectively.
\begin{Definition}[The Transpose of A Vandermonde Matrix]\label{definition_V}
	\begin{equation}
		V(t,[N+1]):=\begin{bmatrix}
			1& 1 &1  &\cdots  &1   \\
			1& 2 &3  &\cdots  &N+1   \\
			1& 2^2 &3^2  &\cdots   &(N+1)^2  \\
			\vdots&\vdots&\vdots&\ddots&\vdots   \\
			1&  2^{t-1}& 3^{t-1} &\cdots& (N+1)^{t-1}
		\end{bmatrix},
	\end{equation}
\end{Definition}
here we let the first column be the sub-matrix $ V_{S} $, the second column correspond to  the first share $ P_1 $ and so on. Due to the determinant of a Vandermonde matrix,  any $ t $ columns corresponding to any $ t $ shares are full rank, so are any $ t-1 $ columns corresponding to any $ t-1 $ shares combined with the first column corresponding to the secret $ S $, then $ V(t,[N+1]) $ is a linear $ (t, N, S) $ threshold scheme.

Recall that from one secret to multiple ones, we can imply the independent combination of $ |\mathcal{T}| $  linear threshold schemes to get a linear $ (N,\mathcal{T}) $ multiple threshold scheme according to corollary \ref{point_simple_complex}. More specifically, given any structure $ (N,\mathcal{T}) $, we prepare a linear $ (t_k,N,S_{k,j}) $ threshold scheme based on $ V(t_k,[N+1]) $ for any $ k\in[K],j\in[|T_k|] $, each of them is also a degradation version of linear $ (N,\mathcal{T}) $ multiple threshold scheme with the strong secure condition. The independent combination of these $ |\mathcal{T}| $ generator matrices leads to a linear $ (N,\mathcal{T}) $ multiple threshold scheme with the strong secure condition such that $ H(S_{k,j})=1,\forall k\in[K],j\in[|T_k|] $ and $ H(P_i)=|\mathcal{T}|,\forall i\in[N] $. And the order of the underlying finite field is still larger than $ N+1 $. Then we have $ \sigma_{N,\mathcal{T}}\leq|\mathcal{T}| $ and $ \tilde{\sigma}_{N,\mathcal{T}}\leq|\mathcal{T}| $. Finally the optimum can be claimed.
\begin{Theorem}
	The infimum of the (average) information ratio equals the number of the secrets, i.e., $ (\tilde{\sigma}_{N,\mathcal{T}})\sigma_{N,\mathcal{T}}=|\mathcal{T}|  $, when under the strong secure condition.
\end{Theorem}
\begin{IEEEproof}
	The proof follows from the above illustrations.
\end{IEEEproof}
\begin{Remark}\normalfont
	The novelty of this theorem compared to the known results of \cite{ramp1999multiple} is embedding the two criteria, i.e., the average information ratio and information ratio. Both criteria show that the more secrets, the less efficiency.
\end{Remark}

\subsection{weak secure}\label{weak_secure_ir}
\subsubsection{converse}
We firstly discuss some converse results that are fundamental for lower bounds of the (average) information ratio when under the weak secure condition.

\begin{Lemma}[Different-Threshold-Bound]\label{Different_Threshold_Bound}
	For any $ (N,\mathcal{T}) $ multiple threshold scheme with the weak secure condition, for any share index $ i\in[N] $ and secret index $ j_k\in[|T_k|] $ of the sub-array $ T_k $, it holds that
	\begin{align}\label{dtb}
		 \sum_{k\in[K]}H(S_{k,j_k})\leq H(P_i).
	\end{align}
\end{Lemma}
\begin{IEEEproof}
	Recall that the difference between the strong secure condition \eqref{strong_condition} and the weak one \eqref{strong_condition} is the knowledge of secrets. More specifically, substitute the secrets $ H(\mathcal{S}_k') $ of the bound  in the proof of lemma \ref{strong_sum_bound} by a single secret $ H(S_{k,j_k}) $, the result follows.
\end{IEEEproof}
\begin{Remark}\normalfont
	Recall that the access structure array $ \mathcal{T} $ can be divided into $ K $ different sub-arrays based on $ K $ unique threshold values. In this way different-threshold-bound says that the sum of the length of different kinds of secrets is no bigger than the length of any secret, which is a weak version of lemma \ref{strong_sum_bound} due to the weak secure condition. This bound is already known as a special case of the result from \cite{herranz2014new}. We find that it is crucial for the (average) information ratio. More specifically, for any structure $ (N,\mathcal{T}) $, the infimum of the information ratio $ \sigma_{N,\mathcal{T}}\geq K $.
\end{Remark}

As mentioned before, we care about every possible structure $ (N,\mathcal{T}) $, and it leads to a discovery of two new bounds. Recall that the access structure array $ \mathcal{T} $ consists of $ K $ sub-arrays, for any $ i\in[K] $ each array $ T_i $ is full of the same threshold value $ t_i $.
\begin{Lemma}[Threshold-Sum-Difference-Bound]\label{Threshold-Sum-Difference-Bound}
	For any $ (N,\mathcal{T}) $ multiple threshold scheme with the weak secure condition, fix an index $ k\in[K] $, similar to lemma \ref{Different_Threshold_Bound} let index $ j_i $ be any element of the set $ [|T_i|] $, also let $ (i_1,\ldots,i_{t_k}) $ be $ t_k $ different values chosen from the set $ [N] $ like lemma \ref{lemma_secret}, then 
	\begin{equation}\label{tsdb}
		t_k\sum_{i\in[k-1]}H(S_{i,j_i})+\sum_{i\in[k:K]}\sum_{j\in[|T_i|]}H(S_{i,j})+\sum_{i\in[k+1:K]}(t_k-t_i)H(S_{i,j_i})\leq\sum_{j\in[t_k]}H(P_{i_j}).
	\end{equation}
\end{Lemma}
\begin{IEEEproof}
	The proof can be divided into three parts, each results in a special bound. The detail is in Appendix \ref{proof_tsdb}.
\end{IEEEproof}
\begin{Remark}\normalfont
	This bound is novel and is important for the (average) information ratio. More specifically, for any structure $ (N,\mathcal{T}) $, the infimum of the information ratio $ \sigma_{N,\mathcal{T}}\geq \max_{k\in[K]}(K-1+|T_k|/t_k+(\sum_{i\in[k+1:K]}|T_i|-t_i)/t_k) $.
\end{Remark}

\begin{Lemma}[Threshold-Product-Bound]\label{Threshold-Product-Bound}
	For any $ (N,\mathcal{T}) $ multiple threshold scheme with the weak secure condition, similar to lemma \ref{lemma_secret} let $ (i_1,\ldots,i_{t_1}) $ be $ t_1 $ different values chosen from the set $ [N] $, then 
	\begin{equation}\label{tpb}
		\sum_{i\in[K]}\frac{\prod_{k\in[K]}t_k}{t_i}\sum_{j\in[|T_i|]}H(S_{i,j})\leq\prod_{k\in[2:K]}t_k\sum_{j\in[t_1]}H(P_{i_j}).
	\end{equation}
\end{Lemma}
\begin{IEEEproof}
	The proof uses induction, where two pattern inequalities are summed at every turn. The detail is in Appendix \ref{proof_tpb}.
\end{IEEEproof}
\begin{Remark}\normalfont
	This bound is novel and is used for the information ratio only. More specifically, for any structure $ (N,\mathcal{T}) $, the infimum of the information ratio $ \sigma_{N,\mathcal{T}}\geq\sum_{i\in[K]}|T_i|/t_i $.
\end{Remark}

We give a lower bound for the average information ratio using different-threshold-bound  and threshold-sum-difference-bound only. Recall that the access structure array $ \mathcal{T} $ consists of $ K $ sub-arrays, for any $ i\in[K] $ each array $ T_i $ of them is full of the same threshold value $ t_i $.
\begin{Lemma}\label{bound_average_information_ratio}
	For any $ (N,\mathcal{T}) $ multiple threshold scheme under the weak secure condition, the average information ratio is closely related to the a unique threshold value of the access structure array $ \mathcal{T} $ and the number of secrets corresponding to it. More specifically, infimum of the average information ratio
	\begin{align}
		\tilde{\sigma}_{N,\mathcal{T}}\geq\frac{|\mathcal{T}|}{\max_{i\in[K]}(\min(t_i,|T_i|))}.
	\end{align}
\end{Lemma}
\begin{IEEEproof}
	The proof uses different-threshold-bound \eqref{dtb} and threshold-sum-difference-bound \eqref{tsdb}. The detail is in Appendix \ref{proof_bound_average_information_ratio}.
\end{IEEEproof}
\begin{Remark}\normalfont
	We will show in the achievability part that this lower bound is tight, which means a specific sub-array $ T_i $ of the access structure array $ \mathcal{T} $ plays a central role for the average information ratio and for a fixed threshold value $ t_i $, a suitable number $ |T_i| $ of secrets improves this performance criteria.
\end{Remark}
	
As for the information ratio, the above three bounds may not be enough to obtain the optimum results, i.e., efforts in investigating more bounds are needed. For example, consider the structure $ (N=3,\mathcal{T}=(3,3,3,3,2,2,2)) $, a new bound is that
\begin{align}\label{future_bound}
	H(S_{1,1})+\sum_{j\in[4]}H(S_{1,j})+2\sum_{j\in[3]}H(S_{2,j})\leq\sum_{i\in[3]}H(P_i)+H(P_3).
\end{align}
\begin{IEEEproof}
	Still rewrite the bound \eqref{tk_plus_one_choose_tk}, we have
	\begin{equation}
		2H(P_{[3]})+H(\mathcal{S}_{2})\leq \sum_{i\in[3]}H(P_{[3]/\{i\}})\nonumber.
	\end{equation}
	Consider that $ H(P_{[3]})=H(\mathcal{S}_{[2]},P_{[3]})\geq H(\mathcal{S}_{[2]}) $ and $ H(P_{[3]})=H(\mathcal{S}_{[2]},P_{[3]})\geq H({S}_{1,1},P_{[2]})=H({S}_{1,1})+H(P_{[2]}) $, the result follows.
\end{IEEEproof}
\begin{Remark}\normalfont
	However, we do not know how to generalize this kind of bound as we care about all possible structure $ (N,\mathcal{T}) $ and whether this new bound plays an important role for the information ratio is unclear. In this way, the infimum of the information ratio is claimed for some special cases only as will be discussed in the achievability part.
\end{Remark}
\subsubsection{achievability}
Then we discuss some linear schemes that are crucial for upper bounds of the (average) information ratio and even the (average) randomness ratio.

Let $ t $ and $ a $ be two positive integers, recall that $ V(t,[a]) $ is a transpose of a Vandermonde matrix from $ \mathbb{F}_q^{t\times a} $, where $ q $ is a prime number larger than $ a $, $ t $ means the number of rows and $ [a] $ is exactly the set of elements of the second row. When under the weak secure condition, we give the following building block for linear multiple threshold schemes:
\begin{Lemma}\label{weak_construction}
	For an $ (N,\mathcal{T}) $ multiple threshold scheme whose access structure array consists of only one unique threshold value, i.e., $ |\mathcal{T}|=|T_1| $. Let $ n=\min(t_1,|T_1|) $, then $ V(t_1,n+N) $ is a linear $ (N,\mathcal{T}) $ multiple threshold scheme under the weak secure condition where each column corresponds to a random variable, e.g., the first column is $ S_{1,1} $, the $ n-th $ one is $ S_{1,n} $, the $ (n+1)-th $ one is $ P_1 $ and so on.
\end{Lemma}
Note that in this way we have $ H(S_{1,i})=H(P_j)=1,\forall i\in[n],j\in[N] $ and $ H(S_{1,i})=0,\forall i\in[n+1:|\mathcal{T}|] $ when the number of secrets is bigger than the threshold value.
\begin{IEEEproof}
	In $ V(t_1,n+N) $, any $ t_1 $ columns are full rank. As $ n\leq t_1 $, $ n $ secrets are statistically independent. The decodable condition follows as the threshold value $ t_1 $ equals the number of rows. Since $ t_1-1+1=t_1 $, the weak secure condition is guaranteed. When $ |\mathcal{T}|>t_1 $, the generator matrix $ V(t_1,t_1+N) $ is actually a linear $ (N,\mathcal{T}') $ multiple threshold scheme where $ \mathcal{T}'\subseteq\mathcal{T} $, from theorem \ref{simple_point_complex} the result follows.
\end{IEEEproof}
\begin{Remark}\normalfont
	Such construction of multiple threshold scheme when under the weak secure condition is already known \cite{karnin1983secret,herranz2014new}.
\end{Remark}

Recall that the access structure array $ \mathcal{T} $ consists of $ K $ sub-arrays, for any $ i\in[K] $ each array $ T_i $ of them is full of the same threshold value $ t_i $. Fix an index $ k\in[K] $ such that $ a=\min(t_k,|T_k|) $ is the maximum among all $ K $ chooses. It follows from lemma \ref{weak_construction} and theorem \ref{simple_point_complex} that $ V(t_k,a+N) $ is a linear $ (N,\mathcal{T}) $ multiple threshold scheme such that $ \sum_{i\in[N]}H(P_i)=N $, $ \sum_{i\in[K]}\sum_{j\in[|T_i|]}H(S_{i,j})=a $, and we have an upper bound of the infimum of the average information ratio $ \tilde{\sigma}_{N,\mathcal{T}}\leq{|\mathcal{T}|}/{\max_{i\in[K]}(\min(t_i,|T_i|))}, $ which matches lemma \ref{bound_average_information_ratio} and the optimum is obtained.
\begin{Theorem}
	The infimum of the average information ratio is decided by a specific access structure sub-array, more specifically, 	
	\begin{align}
		\tilde{\sigma}_{N,\mathcal{T}}=\frac{|\mathcal{T}|}{\max_{i\in[K]}(\min(t_i,|T_i|))},
	\end{align}
when under the weak secure condition.
\end{Theorem}
\begin{IEEEproof}
	The proof follows from the above illustrations.
\end{IEEEproof}
\begin{Remark}\normalfont
	From this criteria of efficiency of a multiple threshold scheme with the weak secure condition, we can learn that for a unique threshold value $ t $, the number of secrets corresponding to $ t $ is best controlled within $ t $. And compared to the strong secure condition, weak one improves the performance by a factor related to the number of secrets, which is also limited by the threshold value.
\end{Remark}

\begin{Theorem}
As for the information ratio, we obtain the optimal results in three cases only:
\begin{enumerate}
	\item When $ |T_i|\leq t_i,\forall i\in[K] $, $ \sigma_{N,\mathcal{T}}=K $.
	\item  When $ |T_i|\geq t_i,\forall i\in[K] $, $ \sigma_{N,\mathcal{T}}=\sum_{i\in[K]}|T_i|/t_i $.
	\item  When there exists only one index $ i\in[K] $ such that $ |T_i|> t_i $, $ \sigma_{N,\mathcal{T}}=\max(K,K-1+|T_k|/t_k+ (\sum_{i\in[k+1:K]}|T_i|-t_i)/t_k)$. 
\end{enumerate}
\end{Theorem}
\begin{IEEEproof}
	The proof follows from the following illustrations.
\end{IEEEproof}

In the first case consider different-threshold-bound \eqref{dtb} only, based on which the derivation of the lower bound for the infimum of the information ratio is carried in the following:
\begin{align}\label{dtb_ratio}
	K\min_{k\in[K],j\in[|T_k|]}H(S_{k,j})\leq\sum_{k\in[K]}H(S_{k,j_k})\leq H(P_i)\leq\max_{i\in[N]}H(P_i),
\end{align}
note that any share index $ i $ is feasible and secret index $ j_k $ is anyone of the set $ [|T_k|] $ from the sub-array $ T_k $.

The number of participants is fixed  to be $ N $, and for every sub-array index $ i\in[K] $, introduce an auxiliary access structure array $ T_i $ extracted from the original access structure array $ \mathcal{T} $, we have $ T_i\subseteq\mathcal{T} $ and $ V(t_i,|T_i|+N) $ is a linear $ (N,T_i) $ multiple threshold scheme and also for the structure $ (N,\mathcal{T}) $ by theorem \ref{simple_point_complex}. As the second and third terms of \eqref{dtb_ratio} from any such linear scheme are equal. Over a unified finite field whose order is sufficiently large, e.g., bigger than $ \max_{i\in[K]}|T_i|+N $, an independent combination of these $ K $ generator matrices leads to a linear $ (N,\mathcal{T}) $ multiple threshold scheme, which facilitates all four terms in \eqref{dtb_ratio} to be equal. 

Finally we have $ \sigma_{N,\mathcal{T}}=K $ when $ |T_i|\leq t_i,\forall i\in[K] $, which means for the information ratio criteria, when the number of secrets corresponding to the same threshold value $ t $ is within $ t $, the number of different threshold values determines the performance.

In the second case we use threshold-product-bound \eqref{tpb} only and apply the same framework above. Similarly for any sub-array index $ i\in[K] $ the generator matrix $ V(t_i,t_i+N) $ is a building block for structure $ (N,{T_i}) $ such that both sides of \eqref{tpb} are equal. The $ K $ underlying finite fields are unified with a prime number bigger than $ \max_{i\in[K]}t_i+N $. Note that the length of every share is equal after any combination since each building block guarantees this property.We still need an independent combination of these $ K $ building blocks to make the length of every secret to be equivalent.  

An example is that for each $ i\in[K] $, consider $ \binom{|T_i|}{t_i} $ different permutations of secrets from the same sub-array $ T_i $, then an initial independent combination is a linear $ (N,T_i) $ multiple threshold scheme $ V_i $ such that $ H(S_{i,j})=\binom{|T_i|-1}{t_i-1},\forall j\in[|T_i|] $ and $ H(P_j)=\binom{|T_i|}{t_i},\forall j\in[N] $. In this way the length of secrets from any sub-array of the access structure array $ \mathcal{T} $ are equal. As for all secrets, for any $ i\in[K] $ we firstly independently combine the same $ V_i $ for $ (\prod_{j\in[K]}\binom{|T_j|-1}{t_j-1})/\binom{|T_i|-1}{t_i-1} $ times to get $ V_i' $, then the final independent combination is carried on $ \{V_1',\ldots,V_K'\} $ directly. 

At last we have $ \sigma_{N,\mathcal{T}}=\sum_{i\in[K]}|T_i|/t_i $ when $ |T_i|\geq t_i,\forall i\in[K] $, which means for the information ratio criteria, when the number of secrets corresponding to the same threshold value $ t $ is bigger than $ t $, the influence is individual for different kinds of threshold values, and the more secrets, the less efficiency. 

Before introducing the last case, a new special linear scheme needs to be illustrated.
\begin{Lemma}\label{B_matrix}
	For the structure $ (N,\mathcal{T}=(T_1,T_2)) $ such that $ |T_1|>t_1>t_2>|T_2| $, there exists a linear multiple threshold scheme $ B(N,T_1,T_2) $ satisfying $ H(S_{1,j})=t_2-|T_2|,\forall j\in[|T_1|] $, $ H(S_{2,j})=|T_1|-t_1,\forall j\in[|T_2|] $ and $ H(P_j)=t_2-|T_2|+|T_1|-t_1,\forall j\in[N] $. The number of rows of $ B(N,T_1,T_2) $ is $ |T_1|t_2-t_1|T_2| $ and the order of the underlying finite field is bigger than $ \max((|T_1|+N)(t_2-|T_2|),N(|T_1|-t_1)) $ and needs to be sufficiently large based on the weak secure condition for secrets in $ \mathcal{T}_2 $ as will be shown in the proof. The specific form of $ B(N,T_1,T_2) $ needs the help of $ V(|T_1|t_2-t_1|T_2|,[(|T_1|+N)(t_2-|T_2|)]) $ and $ V((|T_1|-t_1)t_2,[N(|T_1|-t_1)]) $. The former can be divided into $ |T_1|+N $ sub-matrices each with $ t_2-|T_2| $ columns, then the first $ |T_1| $ ones correspond to secrets from the sub-array $ T_1 $ and the last $ N $ ones are part of the shares. Similarly the latter can be divided into $ N $ sub-matrices each with $ |T_1|-t_1 $ columns, stack the $ i $-th sub-matrix with an all-zero one vertically to be a matrix with $ |T_1|t_2-t_1|T_2| $ rows, and then stack the the former part and this matrix horizontally, we get the sub-matrix corresponding to the $ i $-th share. Finally consider an identity matrix with $ |T_2|(|T_1|-t_1) $ columns, stack it with an all-zero matrix vertically to have $ |T_1|t_2-t_1|T_2| $ rows, then the matrix can be divided into $ |T_2| $ sub-matrices each with $ |T_1|-t_1 $ columns and these matrices correspond to secrets from the sub-array $ T_2 $. For example,
	\begin{equation}\setcounter{MaxMatrixCols}{20}
		B(3,(3,3,3,3),(2))=\begin{bmatrix}
			1& 1 & 1 & 1 & 1 & 1 & 1 & 1 & 1 & 1 & 1  \\
			1&  2&  3&  4&  0&  5&  1&  6&  2&  7&3  \\
			1&  2^2&  3^2&  4^2&  0&  5^2&  0&  6^2&  0&  7^2&0  \\
			1&  2^3&  3^3&  4^3&  0&  5^3&  0&  6^3&  0&  7^3&0  \\
			1&  2^4& 3^4 &  4^4&  0&  5^4&  0&  6^4&  0&  7^4& 0
		\end{bmatrix}.
	\end{equation}
\end{Lemma}
\begin{IEEEproof}
	The proof is carried by checking the system conditions of the definition of linear multiple threshold scheme. The detail can be found in Appendix \ref{proof_B_matrix}.
\end{IEEEproof}
\begin{Remark}\normalfont
	This construction is novel and plays an important role for the information ratio and the (average) randomness ratio.
\end{Remark}

For the third case, both different-threshold-bound \eqref{dtb} and threshold-sum-difference-bound \eqref{tsdb} are considered. In this case assume that $ |T_k|> t_k,k\in[K] $ and $ |T_i|\leq t_i,\forall i\in[K]/\{k\} $, then we combine the two terms of the information ratio with threshold-sum-difference-bound \eqref{tsdb} targeted on $ k $:
	\begin{align}\label{tsdb_ratio}
		&((K-1)t_k+|T_k|+\sum_{i\in[k+1:K]}|T_i|-t_i)\min_{k\in[K],j\in[|T_k|]}H(S_{k,j})\leq\\
	&t_k\sum_{i\in[k-1]}H(S_{i,j_i})+\sum_{i\in[k:K]}\sum_{j\in[|T_i|]}H(S_{i,j})+\sum_{i\in[k+1:K]}(t_k-t_i)H(S_{i,j_i})\\
	&\leq\sum_{j\in[t_k]}H(P_{i_j})\leq t_k\max_{i\in[N]}H(P_i),
\end{align}
the declaration of $ j_i $ and $ i_j $ follows \eqref{tsdb}.

Then we prepare a linear scheme for the structure $ (N,\mathcal{T}) $ by independent combination like above. For any $ i\in[k-1] $ the generator matrix $ V(t_i,|T_i|+N) $ is a building block for structure $ (N,{T_i}) $ such that both sides of \eqref{tpb} or \eqref{tsdb} are equal, so are the building block $ B(N,T_k,T_i),i\in[k+1:K] $ for the structure $ (N,(T_k,T_i)) $ when $ |T_i|<t_i $ and the building block $ V(t_i,[t_i+N]) $ for the structure $ (N,T_i) $ if $ |T_i|=t_i $. As the value of $ \sum_{i\in[k+1:K]}t_i-|T_i| $ minus $ |T_k|-t_k $ cased by the latter $ K-k $ generator matrices may not be zero. We introduce two more pattern building blocks in order to make the length of every secret to be equivalent, i.e., the generator matrix $ V(t_k,t_k+N) $ for the structure $ (N,T_k) $, it only fulfills threshold-sum-difference-bound \eqref{tsdb}, and the generator matrix $ V(t_i,|T_i|+N),i\in[k+1:K] $ for structure $ (N,{T_i}) $ while this only fulfills different-threshold-bound \eqref{dtb} when $ |T_i|<t_i $. 

As in the second case the specific combination coefficients are tedious and by the above introduction of four pattern building blocks we give the conclusion directly. If $ \sum_{i\in[k+1:K]}t_i-|T_i|\geq |T_k|-t_k $, $ \sigma_{N,\mathcal{T}}=K $; otherwise, $ \sigma_{N,\mathcal{T}}=K-1+|T_k|/t_k+ (\sum_{i\in[k+1:K]}|T_i|-t_i)/t_k$. It means for the information ratio criteria, when the circumstances that the number of secrets corresponding to the same threshold value $ t $ is bigger than $ t $ and less than $ t $ both exist, the optimization is complicated and needs further insight.

\section{(Average) randomness ratio}\label{rr}
We firstly discuss multiple threshold scheme with  the strong secure condition, of which some results can be modified for the weak secure condition.
\subsection{strong secure}\label{strong_secure_rr}
We firstly discuss a converse result which is fundamental for lower bounds of the (average) randomness ratio when under the strong secure condition.
\begin{Lemma}\label{strong_threshold_bound}
	For any $ (N,\mathcal{T}) $ multiple threshold scheme with the strong secure condition, it holds that $ \sum_{k\in[K]}\sum_{j\in[|T_k|]}t_kH(S_{k,j})\leq H(P_{[N]}) $.
\end{Lemma}
\begin{IEEEproof}
	The proof is based on the mutual information bound as in lemma \ref{lemma_secret}. The detail is in Appendix \ref{proof_strong_threshold_bound}.
\end{IEEEproof}
\begin{Remark}\normalfont
	This bound says that when under the strong secure condition, the randomness of any $ (N,\mathcal{T}) $ multiple threshold scheme is at least the sum of the product of length of every secret and the corresponding threshold value. This is implied by the work from \cite{ramp1999multiple}, while our proof may be more read-friendly due to the building block in lemma \ref{lemma_secret}. Then consider the (average) randomness ratio, for any $ (N,\mathcal{T}) $ multiple threshold scheme, we apply this bound and it turns out that $ \tau_{N,\mathcal{T}}\geq\sum_{k\in[K]}\sum_{j\in[|T_k|]}(t_k-1) $ and $ \tilde{\tau}_{N,\mathcal{T}}\geq|\mathcal{T}|(t_K-1) $, i.e., lower bounds are formed.
\end{Remark}

As for the achievability part, recall that from one secret to multiple ones, we can imply independent combination of $ |\mathcal{T}| $  linear threshold schemes to get a linear $ (N,\mathcal{T}) $ multiple threshold scheme according to corollary \ref{point_simple_complex}. More specifically, given any structure $ (N,\mathcal{T}) $, we prepare a linear $ (t_k,N,S_{k,j}) $ threshold scheme based on $ V(t_k,[N+1]) $ for any $ k\in[K],j\in[|T_k|] $, each of them is also a degradation version of linear $ (N,\mathcal{T}) $ multiple threshold scheme with the strong secure condition. The independent combination of these $ |\mathcal{T}| $ generator matrices leads to a linear $ (N,\mathcal{T}) $ multiple threshold scheme with the strong secure condition such that $ H(S_{k,j})=1,\forall k\in[K],j\in[|T_k|] $ and $ H(P_{[N]})=\sum_{k\in[K]}\sum_{j\in[|T_k|]}t_k $. Then we have $ \tau_{N,\mathcal{T}}\leq\sum_{k\in[K]}\sum_{j\in[|T_k|]}(t_k-1) $. We also derive an upper bound for the infimum of the average randomness ratio by a single generator matrix $ V(t_K,[N+1]) $, i.e., $ \tilde{\tau}_{N,\mathcal{T}}\leq|\mathcal{T}|(t_K-1) $.  

\begin{Theorem}
	When under the strong secure condition, the infimum of the randomness ratio $ \tau_{N,\mathcal{T}}=\sum_{k\in[K]}|T_k|(t_k-1) $ and the average randomness ratio $ \tilde{\tau}_{N,\mathcal{T}}=|\mathcal{T}|(t_K-1) $.
\end{Theorem}
\begin{IEEEproof}
	The proof follows from the above illustrations.
\end{IEEEproof}
\begin{Remark}\normalfont
	The novelty of this theorem compared to the known results of \cite{ramp1999multiple} is embedding the two criteria, i.e., the average randomness ratio and randomness ratio. Like the average information ratio, a specific threshold value determines the optimal complexity performance by the average randomness ratio criteria. The more secrets, the more complexity.
\end{Remark}

\subsection{weak secure}\label{weak_secure_rr}
\subsubsection{converse}
We firstly discuss some converse results that are fundamental for lower bounds of the (average) randomness ratio when under the weak secure condition.
\begin{Lemma}[Threshold-Value-Bound]\label{All_Threshold_Bound}
	For any $ (N,\mathcal{T}) $ multiple threshold scheme with the weak secure condition, for any secret index $ j_k\in[|T_k|] $ of the sub-array $ T_k $, it holds that
	\begin{align}\label{tvb}
		\sum_{k\in[K]}t_kH(S_{k,j_k})\leq H(P_{[N]}).
	\end{align}
\end{Lemma}
\begin{IEEEproof}
	Recall that the difference between the strong secure condition \eqref{strong_condition} and the weak one \eqref{strong_condition} is the knowledge of secrets. 
	More specifically, substitute the secrets $ H(\mathcal{S}_k') $ of the bound  in the proof of lemma \ref{strong_threshold_bound} by a single secret $ H(S_{k,1}) $, the result follows.
\end{IEEEproof}
\begin{Remark}\normalfont
	Recall that the access structure array $ \mathcal{T} $ can be divided into $ K $ different sub-arrays based on $ K $ unique threshold values. In this way threshold-value-bound says that the randomness of any $ (N,\mathcal{T}) $ multiple threshold scheme is at least the sum of the product of length of different kinds of secret and the corresponding threshold values, which is a weak version of lemma \ref{strong_threshold_bound} due to the weak secure condition. This bound is novel as we investigate randomness under the weak secure condition. We find that it is crucial for the (average) randomness ratio. Consider the $ \prod_{i\in[K]}|T_i| $ different permutations of threshold-value-bound, the sum is equivalent to the following bound
	\begin{equation}\label{permutation_tvb}
		\sum_{k\in[K]}\sum_{j\in[|T_k|]}\frac{t_k}{|T_k|}H(S_{k,j})\leq H(P_{[N]}).
	\end{equation}
\end{Remark}

When under the weak secure condition, there is another novel bound similar to threshold-sum-difference-bound \eqref{tsdb}.
\begin{Lemma}[Threshold-Sum-Bound]\label{Threshold-Sum-Bound}
	For any $ (N,\mathcal{T}) $ multiple threshold scheme with the weak secure condition, fix an index $ k\in[K] $, similar to lemma \ref{Different_Threshold_Bound} let index $ j_i $ be any element of the set $ [|T_i|] $, then 
	\begin{equation}\label{tsb}
		\sum_{i\in[k-1]}t_iH(S_{i,j_i})+\sum_{i\in[k:K]}\sum_{j\in[|T_i|]}H(S_{i,j})\leq H(P_{[N]}).
	\end{equation}
\end{Lemma}
\begin{IEEEproof}
	The proof can be divided into two parts, each results in a special bound. The detail is in Appendix \ref{proof_Threshold-Sum-Bound}.
\end{IEEEproof}
\begin{Remark}\normalfont
	We find that it is crucial for the (average) randomness ratio. Consider the total $ \prod_{i\in[k-1]}|T_i| $ different permutations of threshold-sum-bound, the sum is equivalent to the following bound
	\begin{equation}\label{permutation_tsb}
		\sum_{i\in[k-1]}\sum_{j\in[|T_i|]}\frac{t_i}{|T_i|}H(S_{i,j})+\sum_{i\in[k:K]}\sum_{j\in[|T_i|]}H(S_{i,j})\leq H(P_{[N]}).
	\end{equation}
\end{Remark}

\subsubsection{achievability}
Then we discuss some linear schemes that are crucial for upper bounds of the (average) information ratio.

Recall that linear scheme like $ V(t,[N+1]) $ in definition \ref{definition_V} and lemma \ref{weak_construction} for multiple threshold scheme under the (strong) weak secure condition is already known by the information theory community, and in the last section we propose a new generator matrix $ B(N,T_1,T_2) $ which can be seen as a modified version of $ V(t,[N+1]) $ for specific structure. Here we continue this idea and give the following lemma.

\begin{Lemma}\label{A_matrix}
	For the structure $ (N,\mathcal{T}=(T_1)) $ such that $ |T_1|>t_1 $, there exists a linear multiple threshold scheme $ A(N,T_1,a) $ where integer $ a\in[t_1-1] $ such that $ H(S_{1,j})=a,\forall j\in[|T_1|] $, $ H(P_{i_j})=a,\forall j\in[t_1-a] $ and $ H(P_{i_j})=|T_1|-t_1+a,\forall j\in[t_1-a+1:N] $, here $ (i_1,\ldots,i_{N})$ are $ N $ different values chosen from the set $ [N] $. The number of rows of $ A(N,T_1,a) $ is $ a|T_1| $ and the order of the underlying finite field is bigger than $ \max(a(|T_1|+N),(N-t_1+a)(|T_1|-t_1)) $. The specific form of $ A(N,T_1,a) $ needs the help of $ V(a|T_1|,[a(|T_1|+N)]) $ and $ V(a(|T_1|-t_1),[(N-t_1+a)(|T_1|-t_1)]) $. The former can be divided into $ |T_1|+N $ sub-matrices each with $ a $ columns, then the first $ |T_1| $ ones correspond to secrets from the sub-array $ T_1 $, the next $ t_1-a $ ones for shares $ \{P_{i_1},\ldots,P_{i_{t_1-a}}\} $ and the last $ N-t_1+a $ ones are part of the other $ N-t_1+a $ shares. Similarly the latter can be divided into $ N-t_1+a $ sub-matrices each with $ |T_1|-t_1 $ columns, stack the $ j $-th sub-matrix with an all-zero one vertically to be a matrix with $a|T_1|  $ rows, and then stack the the former part and this matrix horizontally, we get the sub-matrix corresponding to the $ i_{t_1-a+j} $-th share. For example,
	\begin{equation}\setcounter{MaxMatrixCols}{20}
		A(3,(2,2,2),1)=\begin{bmatrix}
			1& 1 & 1 & 			1 &    1 & 		1 & 1 &1  \\
			1&  2&  3&  		4&       5&     0&  6  &0\\
			1&  2^2&  3^2&  4^2&   5^2&  0&  6^2&  0
		\end{bmatrix}.
	\end{equation}
\end{Lemma}
\begin{IEEEproof}
	The proof is carried by checking the system conditions of the definition of linear multiple threshold scheme. The detail can be found in Appendix \ref{proof_A_matrix}.
\end{IEEEproof}
\begin{Remark}\normalfont
	This construction is novel, has different size of shares and plays the extreme direction role \cite{LPbook} for the projection $ \text{proj}_{\mathbf{h}_I}(\Gamma_{N+|T_1|}\cap\mathcal{C}_{0,1,3}) $ when $ t_1<N $. When $ t_1=N $, the cases for $ a\in[2:t_1-1] $ are not extreme directions as the threshold-sum-difference-bound \eqref{tsdb} is unique, while the case that $ a=1 $ is an extreme direction. Recall that for a fixed structure $ (N,T_1) $ with $ |T_1|> t_1 $, in the $(2^{N+|T_1|}-1)$-dimension Euclidean space we denote the coordinates related with $ N $ shares and $ |T_1| $ secrets by $ \mathbf{h}_I $, and the projection is from the polyhedral cone formed by the Shannon-type inequalities and system conditions of the structure $ (N,T_1) $. 
\end{Remark}

Consider the average randomness ratio firstly and there are two cases. Recall that the access structure array $ \mathcal{T} $ consists of $ K $ sub-arrays.

The first one is that for any structure $ (N,\mathcal{T}) $ when there at least exists one index $ i\in[K] $ satisfying $ |T_i|\geq t_i $. We use threshold-sum-bound \eqref{tsb} for $ k=1 $ and have a lower bound of the infimum of the average randomness ratio $ \tilde{\tau}_{N,\mathcal{T}}\geq0 $. Then the corresponding linear scheme can be $ A(N,T_i,1) $ or $ V(t_i,[t_i+N]) $ for the structure $ (N,\mathcal{T}) $ by theorem \ref{simple_point_complex} and it turns out that $ \tilde{\tau}_{N,\mathcal{T}}\leq0 $. The optimum is claimed and means that no additional randomness is needed in this case for the weak secure condition.

The second case is the opposite, that is, for every index $ i\in[K] $, the number of secrets corresponding to the same threshold value $ t_i $ is less than $ t_i $, i.e., $ |T_i|<t_i $. Use the permutation version of threshold-value-bound \eqref{permutation_tvb} we have $ \tilde{\tau}_{N,\mathcal{T}}\geq|\mathcal{T}|\cdot\min_{i\in[K]}(t_i-|T_i|)/|T_i|$. Assume $ k $ is the index matching this minimum value, a linear scheme $ V(t_k,[|T_k|+N]) $ for the structure $ (N,\mathcal{T}) $ leads to the upper bound $ \tilde{\tau}_{N,\mathcal{T}}\leq|\mathcal{T}|(t_k-|T_k|)/|T_k| $. The optimum is claimed and means that under the weak secure condition we can benefit from the number of  secrets corresponding to the same threshold value when consider randomness.
\begin{Theorem}
	For any structure $ (N,\mathcal{T}) $, the infimum of the average randomness ratio
	\begin{align}
		\tilde{\tau}_{N,\mathcal{T}}=|\mathcal{T}|\cdot\max(\min_{i\in[K]}(t_i-|T_i|)/|T_i|,0).
	\end{align}
\end{Theorem}
\begin{IEEEproof}
	The proof follows from the two cases mentioned above.
\end{IEEEproof}

As for the randomness ratio, still two cases will be discussed.

Similarly the first case is  that the number of secrets corresponding to the same threshold value $ t_k $ is bigger than $ t_k $, fix the smallest index from $ [K] $ to be $ k $. Introduce the permutation version of threshold-sum-bound \eqref{permutation_tsb} towards $ k $ and we have a lower bound of the infimum of the randomness ratio $ \tau_{N,\mathcal{T}}\geq\sum_{i\in[k-1]}t_i-|T_i| $. For any index $ i\in[k-1] $ we prepare $ V(t_i,[|T_i|+N]) $ for the structure $ (N,T_i) $, the linear scheme $ A(N,T_k,1) $ is for the structure $ (N,T_k) $ and for every index $ i\in[k+1] $, if $ |T_i|<t_i $, the generator matrix $ B(N,T_k,T_i) $ is used; otherwise, we only prepare $ A(N,T_i,1) $ or $ V(t_i,[t_i+N]) $. Let the order be sufficiently large we have a unified finite field, and the independent combination of these generator matrices leads to a linear $ (N,\mathcal{T}) $ multiple threshold scheme under the weak secure condition such that $ \tau_{N,\mathcal{T}}\leq\sum_{i\in[k-1]}t_i-|T_i| $. Finally the optimum is claimed and we can indeed learn that more secrets help to reduce randomness.

The second case is still the opposite, that is, for any index $ i\in[K] $, $ |T_i|\leq t_i $. Use the permutation version of threshold-value-bound \eqref{permutation_tvb} we have $ {\tau}_{N,\mathcal{T}}\geq\sum_{i\in[K]}t_i-|T_i|$. We prepare a generator matrix $ V(t_i,[|T_i|+N]) $ for every sub-array $ T_i $. Still let the order be sufficiently large, over the unified finite field the independent combination of these $ K $ generator matrices leads to $ \tau_{N,\mathcal{T}}\leq\sum_{i\in[K]}t_i-|T_i| $. The optimum is claimed.

\begin{Theorem}
	For any structure $ (N,\mathcal{T}) $, the infimum of the randomness ratio
	\begin{align}
		{\tau}_{N,\mathcal{T}}=\sum_{i\in[b]}t_i-|T_i|,
	\end{align}
where b depends on the access structure array $ \mathcal{T} $. More specifically, if there exists an index $ k\in[K] $ such that $ |T_k|>t_k $, let $ b $ be the smallest such index minus 1; otherwise, $ b=K $.
\end{Theorem}
\begin{IEEEproof}
	The proof follows from the two cases mentioned above.
\end{IEEEproof}


\section{Conclusions and future direction}\label{conclusion}
\begin{table}\caption{four ratios under two secure conditions}\label{table_asdasd}
	\begin{tabular}{|c|c|c|}
			\hline
			& strong & weak \\
			\hline
			$\tilde{\sigma}_{N,\mathcal{T}}$& $|\mathcal{T}|$ & $ |\mathcal{T}|/\max_{i\in[K]}(\min(t_i,|T_i|)) $ \\
			\hline
			\multirow{4}{*}{${\sigma}_{N,\mathcal{T}}$}& \multirow{4}{*}{$|\mathcal{T}|$} & $K$ \\
			\cline{3-3}
			&  & $ \sum_{i\in[K]}|T_i|/t_i $ \\
			\cline{3-3}
			&  & $ \max(K,K-1+|T_k|/t_k+ (\sum_{i\in[k+1:K]}|T_i|-t_i)/t_k) $ \\
			\cline{3-3}
			&  & Unknown  \\
			
			\hline
			$\tilde{\tau}_{N,\mathcal{T}}$&$ |\mathcal{T}|(t_K-1)$  & $ |\mathcal{T}|\cdot\max(\min_{i\in[K]}(t_i-|T_i|)/|T_i|,0) $  \\
			\hline
			${\tau}_{N,\mathcal{T}}$& $ \sum_{k\in[K]}|T_k|(t_k-1)  $ &$ \sum_{i\in[b]}t_i-|T_i| $  \\
			\hline
		\end{tabular}
\end{table}

In table \ref{table_asdasd} we conclude the results of the four ratios. We can see that when under the weak secure condition, multiple threshold scheme has improved performance of both efficiency and complexity, compared to the strong secure condition.

When under the weak secure condition, we find that these three new bounds come from the case that the number of secrets corresponding to the same threshold value $ t $ is lager than $ t $. As leaded by corollary \ref{bound_complex_simple}, we conjecture that the projection $ \text{proj}_{\mathbf{h}_I}(\Gamma_{N+|\mathcal{T}|}\cap\mathcal{C}_{0,1,3}) $ may have all patterns of bounds, where $ N=4 $, $ \mathcal{T}=(T_1,T_2,T_3) $ with $ t_1=4,|T_1|=5 $, $ t_2=3,|T_2|=4 $ and $ t_3=2,|T_3|=3 $. However, existing numerical experiments may be difficult to carry on due to the tremendous number of the Shannon-type inequalities involved. So the bound  \eqref{future_bound} may also be a good direction to attack.

\appendices

\section{proofs for problem formulation}
\subsection{proof of theorem \ref{subset_region_equal}}\label{proof_theorem_1}
	Let $\mathcal{P}:=\Gamma_{N+|\mathcal{T}'|}\cap\mathcal{C}_{0,1,(2)3}'$ and $\mathcal{Q}:=\text{proj}_{\mathbf{h}_L}(\Gamma_{N+|\mathcal{T}|}\cap\mathcal{C}_{0,1,(2)3}) $ for ease of notation.
	
	$(\Rightarrow)$Form a set of $ |\mathcal{T}|-|\mathcal{T}'| $ excluded secrets $ B:=\{S_j:|\mathcal{T}'|+1\leq j\leq|\mathcal{T}|\} $ and recall that $ \mathcal{O} $ is the set of $ N+|\mathcal{T}| $ random variables. For any vector $ \mathbf{x}\in\mathcal{P} $, there exists a vector $ \mathbf{X}\in\mathcal{H}_{N+|\mathcal{T}|} $ such that $ \mathbf{X}_{\mathbf{h}_L}=\mathbf{x} $, where $ \mathbf{X}_{\mathbf{h}_L} $ is a vector extracted from the original vector $ \mathbf{X} $ according to the coordinates of the region $ \mathcal{P} $. More specifically, for any non-empty $ a\subseteq\mathcal{O}$, let $ X_a=x_{a/B} $, where $ x_\emptyset=0 $ and $ / $ means the difference of two sets, in this way it follows that $ \mathbf{X}_{\mathbf{h}_L}=\mathbf{x} $. The vector $ \mathbf{X} $ satisfies the Shannon-type inequalities since for $ a\subseteq b\subseteq \mathcal{O}$, $ X_b=x_{b/B}\geq x_{a/B}=X_a $ and for $ a,b\subseteq \mathcal{O} $, 
	$ X_a+X_b=x_{a/B}+x_{b/B}\geq x_{a\cup b/B}+x_{a\cap b/B}=X_{a\cup b}+X_{a\cap b} $. $ \mathbf{X}\in\mathcal{C}_0 $ because $ X_{S_1,\ldots,S_{|\mathcal{T}|}}=x_{S_1,\ldots,S_{|\mathcal{T}'|}}=\sum_{i\in[|\mathcal{T}'|]}x_i=\sum_{i\in[|\mathcal{T}|]}X_i$. With the degeneration of the secrets in the set $ B $ from the construction of the vector $ X $ and by the definitions of decodable condition and (strong) weak condition similar to the independent assumption of all secrets, we have $ \mathbf{X}\in\mathcal{C}_{1,(2)3} $. Finally, $ \forall\mathbf{x}\in \mathcal{P}, \mathbf{x}\in \mathcal{Q} $.
	
	$(\Leftarrow)$For any vector $ \mathbf{X}\in\Gamma_{N+|\mathcal{T}|}\cap\mathcal{C}_{0,1,(2)3} $, from which extract a vector $ \mathbf{x} $ according to the coordinates of the region $ \mathcal{P} $. As $ \mathbf{X} $ satisfies the Shannon-type inequalities for $ N+|\mathcal{T}| $ random variables, it follows that $ \mathbf{x}\in\Gamma_{N+|\mathcal{T}'|} $ by the definitions of non-decreasing and submodular properties. With the help of the Shannon-type inequalities, we have $ X_{S_{[|\mathcal{T}'|]}}+\sum_{j=|\mathcal{T}'|+1}^{j=|\mathcal{T}|}X_{S_j}\geq X_{S_{[|\mathcal{T}|]}}=\sum_{j\in[|\mathcal{T}|]}X_{S_j}$ and $ \sum_{j\in[|\mathcal{T}'|]}X_{S_j}\geq X_{S_{[|\mathcal{T}'|]}} $, then $ \mathbf{x}\in\mathcal{C}_0' $. For $ a\subseteq b\subseteq c\subseteq \mathcal{O} $, similarly we have $ X_c-X_a\geq X_b-X_a $ and when $ X_c-X_a=0 $, it follows that $ X_b-X_a=0 $, then $ \mathbf{x}\in\mathcal{C}_1' $. For three disjoint sets $ a,b,c\subseteq \mathcal{O} $, we have $ X_{a,b}+X_{a,c}\geq X_{a,b,c}+X_{c} $ and when $ X_{a,b,c}=X_{a,b}+X_{c} $, it follows that $ X_{a,c}=X_a+X_c $, then if we consider the strong secure condition, $ \mathbf{x}\in\mathcal{C}_2' $. The case of $ \mathcal{C}_3' $ follows for reasons similar to the Shannon-type inequalities for different numbers of random variables above. Hence we have $ \forall\mathbf{x}\in \mathcal{Q}, \mathbf{x}\in \mathcal{P} $.

\subsection{proof of theorem \ref{simple_point_complex}}\label{proof_theorem_2}
		Note that we have assumed all vectors of any sub-matrix $ V_i $ of a generator matrix are linear independent, if $ V_i $ is all-zero, we call such matrix a dummy one since $ \text{rk}(V_i)=0 $. Form a set of $ |\mathcal{T}|-|\mathcal{T}'| $ excluded secrets $ B:=\{S_j:j\in[|\mathcal{T}'|+1,|\mathcal{T}|]\} $. We have already known that the generator matrix $ V $ satisfies the system conditions of the structure $(N,\mathcal{T}')  $, and we will show that after the construction of dummy matrices for the set of excluded secrets $ B $, $ V $ is also for $ (N,\mathcal{T}) $.
		
		For the independent assumption of secrets, $\text{rk}(V_{S_{[|\mathcal{T}|]}})=\text{rk}(V_{S_{[|\mathcal{T}'|]}})=\sum_{i\in[|\mathcal{T}'|]}\text{rk}(V_{S_i})=\sum_{i\in[|\mathcal{T}|]}\text{rk}(V_{S_i})$, the first equation is from the all-zero matrices, the second is by the property of $ V $ and the third is still via the dummy construction. If we have $\text{rk}(V_{{S}_{C},P_A})=\text{rk}(V_{P_A}),A\subseteq[N],C\subseteq[|\mathcal{T}'|]$ under the decodable condition, from the dummy construction it follows that $ \text{rk}(V_{S_B,{S}_{C},P_A})=\text{rk}(V_{{S}_{C},P_A})=\text{rk}(V_{P_A}) $, which is enough for the structure $ (N,\mathcal{T}) $. Similarly consider the (strong) weak secure condition, if $\text{rk}(V_{{S}_{C},P_A})=\text{rk}(V_{P_A})+\text{rk}(V_{S_c}),A\subseteq[N],C\subseteq[|\mathcal{T}'|]$, as the additional matrices are all-zero, $\text{rk}(V_{S_B,{S}_{C},P_A})=\text{rk}(V_{{S}_{C},P_A})=\text{rk}(V_{P_A})+\text{rk}(V_{S_B,S_c})$, which is also sufficient.

\section{proofs for (average) information ratio}
\subsection{proof of lemma \ref{strong_sum_bound}}\label{proof_strong_sum_bound}
	Recall that the access structure array $ \mathcal{T} $ has $ K $ sub-arrays, each of them is denoted by $ T_k $ full of $ t_k $, the same threshold value of the corresponding set of secrets $ \mathcal{S}_k=\{S_{k,1},\ldots,S_{k,|T_k|}\} $. In this way the non-increasing property of $ \mathcal{T} $ says that $ t_1>\cdots>t_K $.
	
	Based on corollary \ref{bound_complex_simple}, we introduce an auxiliary access structure array $ \mathcal{T}'\supseteq\mathcal{T} $. More specifically, $ \mathcal{T}' $ has $ t_1-1 $ sub-arrays with $ t_1'=t_1,t_2'=t_1-1,\ldots,t_{t_1-1}'=2 $, and the length of each is larger than or equal to the corresponding sub-array of $ \mathcal{T} $, i.e., for any $ k\in[t_1-1] $, if there exists $ j\in[K] $ such that $ t_k'=t_j $, then $ |T_k'|=|T_j| $; otherwise, $ |T_k'|=1 $. Then our proof is towards the structure $ (N,\mathcal{T}') $, and the result for the structure $ (N,\mathcal{T}) $ will be obtained from truncation.
	
	For any $ k\in[2:t_1-1] $, from the strong secure condition \eqref{strong_condition}, for all $ A\subseteq[N] $ with $ |A|\leq t_k'-1 $, we have that $ H(\mathcal{S}_k'|P_A)=H(\mathcal{S}_k') $; and from the decodable condition \eqref{decodable_condition}, for all $ B\subseteq[N] $ with $ |B|\geq t_k' $, it holds that $ H(\mathcal{S}_k' |P_B)=0 $. Substitute the single secret $ S $ from lemma \ref{lemma_secret} by the set of secrets $ \mathcal{S}_k' $, we have $$ H(\mathcal{S}_k')\leq I(P_1;P_{t_1-k+2}|P_2,\ldots,P_{t_1-k+1}),  $$where we use the relation that $ t_k'=t_1-k+1 $ due to the continuous nature. 
	
	And the boundary case is that $$ H(\mathcal{S}_1')\leq H(P_1|P_2,\ldots,P_{t_1}). $$ 
	
	Sum these $ t_1-1 $ inequalities together, we have $ \sum_{k\in[t_1-1]}H(\mathcal{S}_k')\leq H(P_1|P_2)\leq H(P_1)$. From the assumption that all secrets all statistically independent, the permutation of the indices of shares as mentioned in the proof of lemma \ref{lemma_secret} and the truncation technique from corollary \ref{bound_complex_simple}, the final result follows.

\subsection{proof of lemma \ref{Threshold-Sum-Difference-Bound}}\label{proof_tsdb}
	Still we use plain number instead of index like $ j_1 $ as in the proof of lemma \ref{lemma_secret} for simplicity, and an auxiliary access structure array $ \mathcal{T}'\supseteq\mathcal{T} $ is introduced like the proof in lemma \ref{strong_sum_bound}. More specifically, the access structure array $ \mathcal{T}' $ consists of $ t_1-1 $ continuous natural numbers from $ t_1 $ to $ 2 $,  we replace the starting index $ 1 $ by $ k+t_k-t_1 $, i.e., $ t_{k+t_k-t_1}'=t_1,\ldots,t_{k+t_k-2}'=2 $, in this way we have $ t_k'=t_k $ from the relation $ t_i'=k+t_k-i $. Fix an index $ k\in[K] $, three parts of inequalities will be presented. 
	
	The first part is about the right hand side of secrets $ \mathcal{S}_k' $, based on the bound of lemma \ref{lemma_secret}, for any index $ i\in[k+1:k+t_k-2] $, we prepare $ i-k $ inequalities in the following:
	\begin{align}
		H(S_{i,1}')&\leq I(P_1;P_{k+t_k-i+1}|P_2,\ldots,P_{k+t_k-i})\nonumber,\\
		H(S_{i,1}')&\leq I(P_2;P_{k+t_k-i+2}|P_3,\ldots,P_{k+t_k-i+1})\nonumber,\\
		&\cdots\nonumber\\
		H(S_{i,1}')&\leq I(P_{i-k};P_{t_k}|P_{i-k+1},\ldots,P_{t_k-1})\nonumber.
	\end{align}
	Then the sum of these $ \sum_{i=k+1}^{k+t_k-2}(i-k) $ inequalities leads to the following bound 
	\begin{equation}
		\sum_{i=k+1}^{k+t_k-2}(i-k)H(S_{i,1}')\leq \sum_{i\in[t_k-2]}H(P_i|P_{i+1})+H(P_{t_k-1},P_{t_k})-H(P_{[t_k]}).
	\end{equation}
	Note that $ t_k'-t_i'=i-k $, $ H(P_i|P_{i+1})\leq H(P_i) $ and $ H(P_{t_k-1},P_{t_k})\leq H(P_{t_k-1})+H(P_{t_k}) $. This concludes the first part and note that if the initial choose $ t_k $ equals the last element of the auxiliary access structure array $ \mathcal{T}' $, i.e., $ 2 $, this part is ignored.
	
	The second part includes the secrets $ \mathcal{S}_k' $ compared to the fist part, i.e., the secrets $ \mathcal{S}_{[k:k+t_k-2]}' $ are considered here. When $ N>t_k $, based on the non-negativeness of conditional mutual information, we list $ t_k $ inequalities in the following:
	\begin{align}
		I(P_1;P_{2},\ldots,P_{t_k+1}|\mathcal{S}_{[k:k+t_k-2]}')&\geq0\nonumber,\\
		I(P_2;P_{3},\ldots,P_{t_k+1}|P_1,\mathcal{S}_{[k:k+t_k-2]}')&\geq0\nonumber,\\
		&\cdots\nonumber\\
		I(P_{t_k};P_{t_k+1}|P_1,\ldots,P_{t_k-1}\mathcal{S}_{[k:k+t_k-2]}')&\geq0\nonumber.
	\end{align}
	Consider the sum of these $ t_k $ inequalities and recall the decodable condition \eqref{decodable_condition}, substitute $ H(P_1,\ldots P_{t_k},\mathcal{S}_{[k:k+t_k-2]}') $ by $ H(P_1,\ldots P_{t_k}) $, for any $ i\in[t_k] $ replace the conditional entropy $ H(P_i|P_{[t_k+1]/\{i\}},\mathcal{S}_{[k:k+t_k-2]}') $ by $ H(P_i|P_{[t_k+1]/\{i\}}) $, and via the independent assumption of secrets \eqref{secrets_independent} we finally have
	\begin{equation}\label{tk_plus_one_choose_tk}
		\sum_{i\in[k:k+t_k-2]}\sum_{j\in|T_i'|}H(S_{i,j}')\leq H(P_{[t_k]})-\sum_{i\in[t_k]}H(P_i|P_{[t_k+1]/\{i\}}).
	\end{equation}
	When $ N=t_k $ we use $ \sum_{i\in[k:k+t_k-2]}\sum_{j\in|T_i'|}H(S_{i,j}')\leq H(P_{[t_k]}) $ by the decodable condition directly.
	
	The third part considers the rest secrets, i.e., $ \mathcal{S}_{[k+t_k-t_1:k-1]}' $. Still based on the bound of lemma \ref{lemma_secret}, for any index $ i\in[k+t_k-t_1+1:k-1] $, $ t_k $ inequalities are prepared in the following:
	\begin{align}
		H(S_{i,1}')&\leq I(P_1;P_{k+t_k-i+1}|P_{[t_k]/\{1\}},P_{t_k+1},\ldots,P_{k+t_k-i})\nonumber,\\
		H(S_{i,1}')&\leq I(P_2;P_{k+t_k-i+1}|P_{[t_k]/\{2\}},P_{t_k+1},\ldots,P_{k+t_k-i})\nonumber,\\
		&\cdots\nonumber\\
		H(S_{i,1}')&\leq I(P_{t_k};P_{k+t_k-i+1}|P_{[t_k]/\{t_k\}},P_{t_k+1},\ldots,P_{k+t_k-i})\nonumber.
	\end{align}
	For the boundary case $ i=k+t_k-t_1 $, we introduce $ t_k $ conditional entropy bounds instead of conditional mutual information ones like the boundary part in the proof of lemma \ref{strong_sum_bound}.
	\begin{align}
		H(S_{k+t_k-t_1,1}')&\leq H(P_1|P_{[t_k]/\{1\}},P_{t_k+1},\ldots,P_{t_1})\nonumber,\\
		H(S_{k+t_k-t_1,1}')&\leq H(P_2|P_{[t_k]/\{2\}},P_{t_k+1},\ldots,P_{t_1})\nonumber,\\
		&\cdots\nonumber\\
		H(S_{k+t_k-t_1,1}')&\leq H(P_{t_k}|P_{[t_k]/\{t_k\}},P_{t_k+1},\ldots,P_{t_1})\nonumber.
	\end{align}
	
	Then the sum of these $ (t_1-t_k)t_k $ inequalities leads to the following bound
	\begin{equation}
		t_k\sum_{i=k+t_k-t_1}^{k-1}H(S_{i,1}')\leq\sum_{i\in[t_k]}H(P_i|P_{[t_k+1]/\{i\}}).
	\end{equation}
	This concludes the third part and note that if the initial choose $ t_k $ equals the first element of the auxiliary access structure array $ \mathcal{T}' $, i.e., $ t_1 $, or $ k=1 $ for short, this part is ignored.
	
	From the permutation of the indices of shares as mentioned in the proof of lemma \ref{lemma_secret} and the truncation technique from corollary \ref{bound_complex_simple}, the final result follows by the sum of these three parts.

\subsection{proof of lemma \ref{Threshold-Product-Bound}}\label{proof_tpb}
	Recall that the access structure array $ \mathcal{T} $ has $ K $ sub-arrays, if $ K=1 $, this bound follows by $ H(\mathcal{S}_1,P_{[t_1]})=H(P_{[t_1]})\geq H(\mathcal{S}_1)$.
	
	When $ K\geq2 $, we introduce an auxiliary access structure array $ \mathcal{T}'\supseteq\mathcal{T} $ as in lemma \ref{strong_sum_bound}. More specifically, $ \mathcal{T}' $ has $ t_1-1 $ sub-arrays with $ t_1'=t_1,t_2'=t_1-1,\ldots,t_{t_1-1}'=2 $, and the length of each is larger than or equal to the corresponding sub-array of $ \mathcal{T} $, i.e., for any $ k\in[t_1-1] $, if there exists $ j\in[K] $ such that $ t_k'=t_j $, then $ |T_k'|=|T_j| $; otherwise, $ |T_k'|=1 $. Then our proof is towards the structure $ (N,\mathcal{T}') $, and the result for the structure $ (N,\mathcal{T}) $ will be obtained from truncation and dividing the same positive integer in both sides of the inequality.
	
	For every $ k\in[2:t_1-1] $, we prepare a pattern inequality which is rewritten from the bound \eqref{tk_plus_one_choose_tk}, one of the three ingredients for the threshold-sum-difference-bound:
	\begin{equation}
		t_k'H(P_{[t_k'+1]})+H(\mathcal{S}_{[k:t_1-1]}')\leq \sum_{i\in[t_k'+1]}H(P_{[t_k'+1]/\{i\}}).
	\end{equation}
	
	The core idea is that special permutations of every pattern inequality except the first one are summed to cancel the entropies of shares. More specifically, the term $ H(P_{[t_k'+1]}) $ can be permuted to cancel the term $ \sum_{i\in[t_{k-1}'+1]}H(P_{[t_{k-1}'+1]/\{i\}}) $ of a new pattern inequality ahead, also this pattern inequality for $ k-1 $ can be multiplied to maintain integral coefficients. 
	Such induction procedure leads to
	\begin{align}
		\prod_{k\in[2:t_1-1]}kH(P_{[t_1]})+\sum_{i\in[2:t_1-1]}\frac{\prod_{k\in[2:t_1]}k}{t_{i-1}t_i}H(\mathcal{S}_{[i:t_1-1]}')\leq\prod_{k\in[2:t_1-1]}k\sum_{i\in[t_1]}H(P_i).
	\end{align}
	Note that similar to the case $ K=1 $, $ H(P_{[t_1]})\geq H(\mathcal{S}_{[t_1-1]}') $, due to the assumption of mutually independent secrets, the permutation of indices of $ N $ shares and the truncation technique from corollary \ref{bound_complex_simple}, the final result follows.

\subsection{proof of lemma \ref{bound_average_information_ratio}}\label{proof_bound_average_information_ratio}

	The proof is carried via case analysis where two cases are involved.
	
	The first case is that for every $ i\in[K] $, the length of the sub-array $ T_i $ is not bigger than the unique threshold value $ t_i $ of it, i.e., $ |T_i|\leq t_i $. Assume $ \max_{i\in[K]}(|T_i|)=|T_k| $, then we introduce an auxiliary access structure array $ \mathcal{T}' $ such that $ \forall i\in[K],|T_i'|=|T_k|,t_i'=t_i $. Note that the number of all different different-threshold-bounds \eqref{dtb} is $ N|T_k|^{K} $, the sum of them is the following:
	\begin{align}
		N|T_k|^{K-1}\sum_{i\in[K]}\sum_{j\in[|T_k|]}H(S_{i,j}')\leq |T_k|^{K}\sum_{i\in[N]}H(P_i).\nonumber
	\end{align}
	Based on the technique of truncation as in corollary \ref{bound_complex_simple} we have $ \tilde{\sigma}_{N,\mathcal{T}}\geq{|\mathcal{T}|}/{|T_k|} $.
	
	Another case is the opposite, i.e., there at least exists one sub-array such that the length of it is strictly bigger than the corresponding threshold value, assume the first one is $ T_k $ and then $ |T_i|\leq t_i,i\in[k-1] $ where $ [0] $ is the empty set. In this way let $ a=\max(|T_1|,\ldots,|T_k-1|,t_k) $ and due to the non-increasing assumption of the access structure array it follows that for any $ i\in[k+1:K] $, $ a\geq\min(t_i,|T_i|) $. The auxiliary access structure array $ \mathcal{T}' $ is constructed similarly,  $ |T_i'|=a,i\in[k-1],|T_i'|=|T_i|,i\in[k:K] $ and the $ K $ unique threshold values are the same. We only use the first two parts in the left  hand side of threshold-sum-difference-bound \eqref{tsdb}, i.e., such relaxation ignores the difference part. Consider the permutation of secrets firstly, the sum of $ a^{k-1} $ such inequalities for fixed $ t_k $ different shares is the following:
	\begin{align}
		t_ka^{k-2}\sum_{i\in[k-1]}\sum_{j\in[|T_i|]}H(S_{i,j}')+a^{k-1}\sum_{i\in[k:K]}\sum_{j\in[|T_i|]}H(S_{i,j}')\leq a^{k-1}\sum_{i\in[t_k]}H(P_i),\nonumber
	\end{align}
	note that when $ k=1 $, the first part in the left hand side is ignored and $ a^{k-1}=t_ka^{k-2} $ in the second part; if $ k\geq 2 $, we relax the coefficient $ a^{k-1} $ of the second part to be $ t_ka^{k-2} $. Finally consider the permutation of $ N $ shares, the sum is the following:
	\begin{align}
		\binom{N}{t_k}t_ka^{k-2}\sum_{i\in[K]}\sum_{j\in[|T_i|]}H(S_{i,j}')\leq \binom{N-1}{t_k-1}a^{k-1}\sum_{i\in[K]}H(P_i).\nonumber
	\end{align}
	Still based on the truncation we have $ \tilde{\sigma}_{N,\mathcal{T}}\geq{|\mathcal{T}|}/{a} $.

\subsection{proof of lemma \ref{B_matrix}}\label{proof_B_matrix}
	Note that in a Vandermonde matrix any sub-matrix is full rank. Denote $ B(N,T_1,T_2) $ by $ V $ for short. 
	
	For all secrets we have $ \text{rk}(V_{\mathcal{S}_{[2]}})=|T_2|(|T_1|-t_1)+|T_1|(t_2-|T_2|)= |T_1|t_2-t_1|T_2|$ as the identity matrix in the upper right corner of $ V_{\mathcal{S}_{[2]}} $ is full rank, so is the sub-matrix in the lower left corner since it is the sub-matrix of a Vandermonde matrix, and that the lower right corner is all-zero facilitates applying Gaussian-Elimination to obtain rank.
	
	For any $ i\in[N] $ split the matrix corresponding to share $ P_i $ into two parts, that is, let $ V_{P_i}^l $ be the first $ t_2-|T_2| $ columns and the last $ |T_1|-t_1 $ columns form $ V_{P_i}^r $. 
	
	As for decodable condition, firstly note that the rank of the matrix formed by any $ t_1 $ shares like $ V_{P_{[t_1]}} $ equals the number of rows of $ V $. The reason is that the upper $ t_2(|T_1|-t_1) $ rows of the matrix $ V_{P_{[t_1]}}^r $, which is the concatenation of the right part of every $ V_{P_i},i\in[t_1] $ horizontally, is full rank as the number of columns is bigger than that of rows, then consider the lower $ t_1(t_2-|T_2|) $ rows of the matrix $ V_{P_{[t_1]}}^l $, it is full rank and such lower rows of $ V_{P_{[t_1]}}^r $ are all-zero. In this way we have $ \text{rk}(V_{P_{[t_1]}})=t_2(|T_1|-t_1)+t_1(t_2-|T_2|)= |T_1|t_2-t_1|T_2|$, which is also the value of $ \text{rk}(V_{\mathcal{S}_{[2]},P_{[t_1]}}) $. Secondly consider the matrix formed by any $ t_2 $ shares like $ V_{P_{[t_2]}} $, of which the upper $ t_2(|T_1|-t_1) $ rows  of the matrix $ V_{P_{[t_2]}}^r $ are full rank and the other rows are all-zero. The upper $ |T_2|(|T_1|-t_1) $ rows of the matrix corresponding to the secrets $ \mathcal{S}_2 $ are full rank and the other rows are all-zero. As $ t_2>|T_2| $, then every column of $ V_{\mathcal{S}_2} $ equals a linear combination of the columns from $ V_{P_{[t_2]}}^r $, we have  $ \text{rk}(V_{P_{[t_2]}})=\text{rk}(V_{\mathcal{S}_2,P_{[t_2]}})$.
	
	At last consider weak secure condition. The rank of the matrix formed by any $ t_1-1 $ shares like $ V_{P_{[t_1-1]}} $ equals $ (t_1-1)(t_2-|T_2|)+t_2(|T_1|-t_1) $ since both $ V_{P_{[t_1-1]}}^r $ and lower $ t_1(t_2-|T_2|) $ rows of $ V_{P_{[t_1-1]}}^l $ are full rank, the lower $ t_1(t_2-|T_2|) $ all-zeros rows of $ V_{P_{[t_1-1]}}^r $ facilitate this observation by Gaussian-Elimination. Stack the matrix corresponding to any secret $ S_{1,j},j\in[|T_1|] $ and $ V_{P_{[t_1-1]}} $ horizontally, note that the former is extracted from $ V(|T_1|t_2-t_1|T_2|,[(|T_1|+N)(t_2-|T_2|)]) $, similarly we have $ \text{rk}(V_{S_{1,j},P_{[t_1-1]}})=\text{rk}(V_{S_{1,j}})+\text{rk}(V_{P_{[t_1-1]}})=|T_1|t_2-t_1|T_2| $. Then we need to focus on the matrix formed by any $ t_2-1 $ shares like $ V_{P_{[t_2-1]}} $. The  $ \text{rk}(V_{S_{1,j},P_{[t_2-1]}})=\text{rk}(V_{S_{1,j}})+\text{rk}(V_{P_{[t_2-1]}})=t_2(t_2-|T_2|)+(t_2-1)(|T_1|-t_1) $ part follows the same procedure as for $ t_1-1 $ shares. 
	
	The case for any secret $ S_{2,j},j\in[|T_2|] $ is not straightforward like before. Recall the matrix $ V_{S_{2,j}} $ corresponding to secret $ S_{2,j} $, it has an identity matrix embedded between the $ (j-1)(|T_1|-t_1)+1 $-th row and the $ j(|T_1|-t_1) $-th row, the other elements of $ V_{S_{2,j}} $ are all-zero. Stack it with $ V_{P_{[t_2-1]}}^r $ horizontally, for weak secure condition we only need to consider the rank of a special matrix extracted from $ V_{P_{[t_2-1]}}^r $, i.e., the concatenation of the first $ (j-1)(|T_1|-t_1) $ rows and $ (t_2-j)(|T_1|-t_1) $ rows between the $ j(|T_1|-t_1)+1 $-th row and the $ t_2(|T_1|-t_1) $-th row. The determinant of this matrix is related to Schur polynomials \cite{prasad2018introduction} which is not $ 0 $ when the order of the underlying finite  field is sufficiently large. For example we can calculate the specific determinants in real number field for all cases, any of which is the product of the sum of monomials and Vandermonde determinant, thus bigger than zero, finally choose the finite field order to be larger than any determinant. In this way  $ \text{rk}(V_{S_{2,j},P_{[t_2-1]}})=\text{rk}(V_{S_{2,j}})+\text{rk}(V_{P_{[t_2-1]}})=(t_2-1)(t_2-|T_2|)+t_2(|T_1|-t_1) $.

\section{proofs for (average) randomness ratio}
\subsection{proof of lemma \ref{strong_threshold_bound}}\label{proof_strong_threshold_bound}

	Recall that the access structure array $ \mathcal{T} $ has $ K $ sub-arrays, each of them is denoted by $ T_k $ full of $ t_k $, the same threshold value of the corresponding set of secrets $ \mathcal{S}_k=\{S_{k,1},\ldots,S_{k,|T_k|}\} $. In this way the non-increasing property of $ \mathcal{T} $ says that $ t_1>\cdots>t_K $.
	
	Based on corollary \ref{bound_complex_simple}, we introduce an auxiliary access structure array $ \mathcal{T}'\supseteq\mathcal{T} $. More specifically, $ \mathcal{T}' $ has $ t_1-1 $ sub-arrays with $ t_1'=t_1,t_2'=t_1-1,\ldots,t_{t_1-1}'=2 $, and the length of each is larger than or equal to the corresponding sub-array of $ \mathcal{T} $, i.e., for any $ k\in[t_1-1] $, if there exists $ j\in[K] $ such that $ t_k'=t_j $, then $ |T_k'|=|T_j| $; otherwise, $ |T_k'|=1 $. Then our proof is towards the structure $ (N,\mathcal{T}') $, and the result for the structure $ (N,\mathcal{T}) $ will be obtained from truncation.
	
	If $ K\geq2 $, for any $ k\in[2:t_1-1] $, from the strong secure condition \eqref{strong_condition}, for all $ A\subseteq[N] $ with $ |A|\leq t_k'-1 $, we have that $ H(\mathcal{S}_k'|P_A)=H(\mathcal{S}_k') $; and from the decodable condition \eqref{decodable_condition}, for all $ B\subseteq[N] $ with $ |B|\geq t_k' $, it holds that $ H(\mathcal{S}_k' |P_B)=0 $. Substitute the single secret $ S $ from lemma \ref{lemma_secret} by the set of secrets $ \mathcal{S}_k' $, for any index $ i\in[t_1-k+1] $ where we use the relation that $ t_k'=t_1-k+1 $ due to the continuous nature, $ t_1-k+1 $ inequalities are prepared in the following:
	\begin{align}
		H(\mathcal{S}_k')&\leq I(P_1;P_{t_1-k+2}|P_{[t_1-k+1]/\{1\}})\nonumber,\\
		H(\mathcal{S}_k')&\leq I(P_2;P_{t_1-k+2}|P_{[t_1-k+1]/\{2\}})\nonumber,\\
		&\cdots\nonumber\\
		H(\mathcal{S}_k')&\leq I(P_{t_1-k+1};P_{t_1-k+2}|P_{[t_1-k+1]/\{t_1-k+1\}})\nonumber.
	\end{align}
	For the boundary case $ k=1 $, we introduce $ t_1 $ conditional entropy bounds instead of conditional mutual information ones like in the proof of lemma \ref{strong_sum_bound}.
	\begin{align}
		H(\mathcal{S}_1')&\leq H(P_1|P_{[t_1]/\{1\}})\nonumber,\\
		H(\mathcal{S}_1')&\leq H(P_2|P_{[t_1]/\{2\}})\nonumber,\\
		&\cdots\nonumber\\
		H(\mathcal{S}_1')&\leq H(P_{t_1}|P_{[t_1]/\{t_1\}})\nonumber.
	\end{align}
	Then the sum of these $ (t_1-1)(t_1+2)/2 $ inequalities leads to the following bound
	\begin{equation}
		\sum_{k\in[K]}t_k'H(\mathcal{S}_k')\leq H(P_{[t_1]})-I(P_1;P_2)\leq H(P_{[N]}).
	\end{equation}
	
	If $ K=1 $, similarly consider the above $ t_1 $ conditional entropy bounds  and the second part in the proof of threshold-sum-difference-bound \eqref{tsdb}. The sum leads to
	\begin{equation}
		t_1H(\mathcal{S}_1')\leq H(P_{[t_1]})-\sum_{k\in[t_1-1]}I(P_k;P_{[k+1:t_1]}|P_{[k-1]})\leq H(P_{[N]}).
	\end{equation}

\subsection{proof of lemma \ref{Threshold-Sum-Bound}}\label{proof_Threshold-Sum-Bound}

	Still we use plain number instead of index like $ j_1 $ as in the proof of lemma \ref{lemma_secret} for simplicity, and an auxiliary access structure array $ \mathcal{T}'\supseteq\mathcal{T} $ is introduced like the proof in lemma \ref{strong_sum_bound}. More specifically, the access structure array $ \mathcal{T}' $ consists of $ t_1-1 $ continuous natural numbers from $ t_1 $ to $ 2 $,  we replace the starting index $ 1 $ by $ k+t_k-t_1 $, i.e., $ t_{k+t_k-t_1}'=t_1,\ldots,t_{k+t_k-2}'=2 $, in this way we have $ t_k'=t_k $ from the relation $ t_i'=k+t_k-i $. 
	
	When $ k=1 $ we use $ \sum_{i\in[k+t_k-t_1:k+t_k-2]}\sum_{j\in|T_i'|}H(S_{i,j}')\leq H(P_{[N]}) $ by decodable condition directly. Otherwise fix an index $ k\in[2:K] $, two parts of inequalities will be presented. 
	
	The first part considers the secrets $ \mathcal{S}_{[k:k+t_k-2]}' $. Based on the non-negativeness of conditional mutual information, we list $ t_k $ inequalities in the following:
	\begin{align}
		I(P_1;P_{2},\ldots,P_{t_k+1}|\mathcal{S}_{[k:k+t_k-2]}')&\geq0\nonumber,\\
		I(P_2;P_{3},\ldots,P_{t_k+1}|P_1,\mathcal{S}_{[k:k+t_k-2]}')&\geq0\nonumber,\\
		&\cdots\nonumber\\
		I(P_{t_k};P_{t_k+1}|P_1,\ldots,P_{t_k-1}\mathcal{S}_{[k:k+t_k-2]}')&\geq0\nonumber.
	\end{align}
	Consider the sum of these $ t_k $ inequalities and recall the decodable condition \eqref{decodable_condition}, substitute $ H(P_1,\ldots P_{t_k},\mathcal{S}_{[k:k+t_k-2]}') $ by $ H(P_1,\ldots P_{t_k}) $, for any $ i\in[t_k] $ replace the conditional entropy $ H(P_i|P_{[t_k+1]/\{i\}},\mathcal{S}_{[k:k+t_k-2]}') $ by $ H(P_i|P_{[t_k+1]/\{i\}}) $, and via the independent assumption of secrets \eqref{secrets_independent} we finally have
	\begin{equation}
		\sum_{i\in[k:k+t_k-2]}\sum_{j\in|T_i'|}H(S_{i,j}')\leq H(P_{[t_k]})-\sum_{i\in[t_k]}H(P_i|P_{[t_k+1]/\{i\}}).
	\end{equation}
	
	The second part considers the rest secrets, i.e., $ \mathcal{S}_{[k+t_k-t_1:k-1]}' $. Still based on the bound of lemma \ref{lemma_secret}, for any index $ i\in[k+t_k-t_1+1:k-1] $, $ t_i'=k+t_k-i $ inequalities are prepared in the following:
	\begin{align}
		H(S_{i,1}')&\leq I(P_1;P_{k+t_k-i+1}|P_{[k+t_k-i]/\{1\}})\nonumber,\\
		H(S_{i,1}')&\leq I(P_2;P_{k+t_k-i+1}|P_{[k+t_k-i]/\{2\}})\nonumber,\\
		&\cdots\nonumber\\
		H(S_{i,1}')&\leq I(P_{k+t_k-i};P_{k+t_k-i+1}|P_{[k+t_k-i]/\{k+t_k-i\}})\nonumber.
	\end{align}
	For the boundary case $ i=k+t_k-t_1 $, we introduce $ t_1 $ conditional entropy bounds instead of conditional mutual information ones like in the proof of lemma \ref{lemma_secret}.
	\begin{align}
		H(S_{k+t_k-t_1,1}')&\leq H(P_1|P_{[t_1]/\{1\}})\nonumber,\\
		H(S_{k+t_k-t_1,1}')&\leq H(P_2|P_{[t_1]/\{2\}})\nonumber,\\
		&\cdots\nonumber\\
		H(S_{k+t_k-t_1,1}')&\leq H(P_{t_1}|P_{[t_1]/\{t_1\}})\nonumber.
	\end{align}
	Then the sum of these $ (t_1-t_k)(t_1+t_k+1)/2 $ inequalities leads to the following bound
	\begin{equation}
		\sum_{i=k+t_k-t_1}^{k-1}(k+t_k-i)H(S_{i,1}')\leq-H(P_{[t_k]})+\sum_{i\in[t_k]}H(P_i|P_{[t_k+1]/\{i\}})+H(P_{[t_1]}).
	\end{equation}
	
	From the permutation of the indices of shares as mentioned in the proof of lemma \ref{lemma_secret} and the truncation technique from corollary \ref{bound_complex_simple}, the final result follows by the sum of these two parts.

\subsection{proof of lemma \ref{A_matrix}}\label{proof_A_matrix}

	Note that in a Vandermonde matrix any sub-matrix is full rank. Denote $ A(N,T_1,a) $ by $ V $ for short. 
	
	For all secrets we have $ \text{rk}(V_{\mathcal{S}_{1}})=a|T_1|$ since it is the sub-matrix of a Vandermonde matrix.
	
	For any $ t_1 $ shares, we have $ at_1 $ different columns extracted from $ V(a|T_1|,[a(|T_1|+N)]) $ and at least $ a(|T_1|-t_1) $ different columns from $ V(a(|T_1|-t_1),[(N-t_1+a)(|T_1|-t_1)]) $ stacked with an all-zero matrix. From this view stack these two matrices horizontally, we can see that the upper right corner is full rank, so is the sub-matrix in the lower left corner since both are the sub-matrices of Vandermonde matrices, and that the lower right corner is all-zero facilitates applying Gaussian-Elimination to obtain the decodable condition, i.e., $ \text{rk}(V_{P_{d}})=\text{rk}(V_{\mathcal{S}_{1},P_{d}})=a|T_1| $ where $ P_d $ denotes any $ t_1 $ shares.
	
	For any $ t_1-1 $ shares, we have $ a(t_1-1) $ different columns extracted from $ V(a|T_1|,[a(|T_1|+N)]) $, regardless of the number of participants $ N $ we assume there are at most $ (t_1-1)(|T_1|-t_1) $ different columns from $ V(a(|T_1|-t_1),[(|T_1|-t_1)\cdot\max(N-t_1+a,t_1-1)]) $ stacked with an all-zero matrix. From this view stack these two matrices horizontally and denote any $ t_1-1 $ shares by $ P_s $ we can see that $ \text{rk}(V_{P_{s}})=a(t_1-1)+a(|T_1|-t_1)=a(|T_1|-1) $ as $ a\leq t_1-1 $ and following the similar procedure in decodable condition above. Then consider any secret, we have $ a $ different columns extracted from $ V(a|T_1|,[a(|T_1|+N)]) $, stack these two matrices horizontally and the final matrix is still full rank. Finally we have $ \text{rk}(V_{S_{1,j},P_{s}})=\text{rk}(V_{S_{1,j}})+\text{rk}(V_{P_{s}})=a|T_1|,j\in[|T_1|] $.
	
	When $ a=1 $ we find that any $ |T_1|-t_1 $ columns from $ V(a(|T_1|-t_1),[(N-t_1+a)(|T_1|-t_1)]) $ form an square matrix which is full rank. Then this part can be substituted by an identity matrix and the order of the underlying finite field may be smaller.

\bibliographystyle{unsrt}
\bibliography{ref}
\end{document}